\documentclass[10pt]{emulateapj}
\usepackage{apjfonts}
\usepackage{graphicx}
\usepackage{psfig}

\newcommand{\begit}{\begin{itemize}}
\newcommand{\enit}{\end{itemize}}
\newcommand{\begen}{\begin{enumerate}}
\newcommand{\enen}{\end{enumerate}}

\newcommand       \be           {\begin{equation}}
\newcommand       \ee           {\end{equation}}
\newcommand       \bea          {\begin{eqnarray}}
\newcommand       \eea          {\end{eqnarray}}

\newcommand       \kms		{\,{\rm km \,\, s}^{-1}}
\newcommand       \cm		{\,{\rm cm }}
\newcommand       \pc		{\,{\rm pc }}
\newcommand       \yr		{\,{\rm yr }}
\newcommand       \s		{\,{\rm s }}
\newcommand       \dynes	{\,{\rm dynes }}
\newcommand       \g		{\,{\rm g }}
\newcommand       \kpc		{\,{\rm kpc }}
\newcommand       \K		{\,{\rm K }}

\newcommand       \KeV		{\,{\rm KeV }}
\newcommand       \erg		{\,{\rm erg }}
\newcommand       \ergs		{\,{\rm erg \,\, s}^{-1}}
\newcommand       \Myr		{\,{\rm Myr} }
\newcommand       \rcl		{r_{\rm cl}}
\newcommand       \mcl		{M_{\rm cl}}
\newcommand       \ecl		{\epsilon_{\rm cl}}
\newcommand       \Rg		{R_{\rm GMC}}
\newcommand       \Mg		{M_{\rm GMC}}
\newcommand       \eg		{\epsilon_{\rm GMC}}
\newcommand{\beqa}{\begin{eqnarray}} 
\newcommand{\eeqa}{\end{eqnarray}}

\begin{document}

\title{THE DISRUPTION OF GIANT MOLECULAR CLOUDS BY RADIATION PRESSURE  \\ \&
THE EFFICIENCY OF STAR FORMATION IN GALAXIES}

\author{Norman Murray\altaffilmark{1,2}, 
Eliot Quataert\altaffilmark{3}, \& Todd A.~Thompson\altaffilmark{4,5,6}}

\altaffiltext{1}{Canada Research Chair in Astrophysics}
\altaffiltext{2}{Canadian Institute for Theoretical Astrophysics, 60
St.~George Street, University of Toronto, Toronto, ON M5S 3H8, Canada;
murray@cita.utoronto.ca} 
\altaffiltext{3}{Astronomy Department \& Theoretical Astrophysics
Center, 601 Campbell Hall, The University of California, Berkeley, CA
94720; eliot@astro.berkeley.edu} 
\altaffiltext{4}{Department of Astronomy, The Ohio State University,
  140 W 18th Ave., Columbus, OH 43210; thompson@astronomy.ohio-state.edu.}
\altaffiltext{5}{Center for Cosmology \& Astro-Particle Physics, The
  Ohio State University, 191 W. Woodruff Ave., Columbus, OH 43210}
\altaffiltext{6}{Alfred P.~Sloan Fellow}

\begin{abstract}
  Star formation is slow, in the sense that the gas consumption time
  is much longer than the dynamical time. It is also inefficient;
  essentially all star formation in local galaxies takes place in
  giant molecular clouds (GMCs), but the fraction of a GMC converted
  to stars is very small, $\sim5\%$. While there is some disagreement
  over the lifespan of GMCs, there is a consensus that it is no more
  than a few cloud dynamical times. In the most luminous starbursts,
  the GMC lifetime is shorter than the main sequence lifetime of even
  the most massive stars, so that supernovae can play no role in GMC
  disruption; another feedback mechanism must dominate.  We
  investigate the disruption of GMCs across a wide range of galaxies,
  from normal spirals to the densest starbursts; we take into account
  the effects of HII gas pressure, shocked stellar winds, protostellar
  jets, and radiation pressure produced by the absorption and
  scattering of starlight on dust grains. In the Milky Way, we find
  that a combination of three mechanisms --- jets, HII gas pressure,
  and radiation pressure --- disrupts the clouds.  In more rapidly
  star forming galaxies such as ``clump'' galaxies at high-redshift,
  ultra-luminous infrared galaxies (ULIRGs) and submillimeter
  galaxies, radiation pressure dominates natal cloud distribution.  We
  predict the presence of $\sim10-20$ clusters with masses
  $\sim10^7M_\odot$ in local ULIRGs such as Arp 220 and a similar
  number of clusters with $M_*\sim10^8M_\odot$ in high redshift clump
  galaxies; submillimeter galaxies will have even more massive
  clusters. We find that the mass fraction of a GMC that ends up in
  stars is an increasing function of the gas surface density of a
  galaxy, reaching $\sim 35\%$ in the most luminous starbursts.
  Furthermore, the disruption of bubbles by radiation pressure stirs
  the interstellar medium to velocities of $\sim10\kms$ in normal
  galaxies and to $\sim100\kms$ in ULIRGs like Arp 220, consistent
  with observations.  Thus, radiation pressure may play a dominant
  role in the ISM of star-forming galaxies.
\end{abstract}


\section{INTRODUCTION}
\label{sec:intro}
The Kennicutt law (Kennicutt 1998)
\be \label{eqn: Kennicutt}%
\dot \Sigma_* =\eta\Sigma_g \Omega
\ee %
relates the star formation surface density $\dot\Sigma_*$ to the gas surface
density $\Sigma_g$ and the local dynamical time $\Omega\approx v_c/R$
in disk galaxies, where $v_c$ is the circular velocity of the galaxy 
and $R$ is the distance from the galactic center. The dimensionless
constant $\eta\approx 0.017$ is surprisingly small, a finding that is
interpreted as showing that star formation is a slow process. 
Star formation is similarly slow on smaller scales within
galaxies (Kennicutt et al.~2007; Bigiel et al.~2008;
Leroy et al.~2008; Krumholz \& Tan 2007).

Remarkably, equation (\ref{eqn: Kennicutt})
holds for galaxies like the Milky Way, with rather
modest star formation rates of order a solar mass per year, for
starburst galaxies with star formation rates of order tens of solar
masses per year, for ultraluminous infrared galaxies (ULIRGs) with
star formation rates around one hundred solar masses per year, and for
sub-millimeter galaxies with star formation rates in excess of one
thousand solar masses per year.  There are indications, however, that
the star formation efficiency $\eta$ may be larger in ULIRGs and
sub-mm galaxies, with $\eta \sim 0.1$ \citep{bouche}.

The large range of galaxies that obey equation (\ref{eqn: Kennicutt})
suggests that whatever process sets the efficiency of star formation
operates in galaxies with very different conditions in their
interstellar media. For example, the gas surface density in the Milky
Way at $8\kpc$ is $\Sigma_g\approx2\times10^{-3}\g\cm^{-2}$
\citep{boulares}, while that in the two $100\pc$ star forming disks of
the ultra-luminous infrared galaxy (ULIRG) Arp 220 is
$\Sigma_g\approx7\g\cm^{-2}$; the mean gas densities of the two
galaxy's star forming disks also differ by a factor of $\gtrsim10^4$.
Although the range in turbulent velocities in the ISM is not so
dramatic, from $\sim6\kms$ in the Milky Way to $\sim60-80\kms$ in Arp
220 \citep{downes}, the turbulent pressure in Arp 220 exceeds that in
the Galaxy by a factor of $\sim10^6$.

Another signature of inefficient star formation relates to individual
giant molecular clouds (GMCs). In the Milky Way, all stars are
believed to form in such clouds. However, the fraction $\eg$ of a GMC
that is turned into stars is quite low, around $5\%$ in the Milky Way
\citep{1997ApJ...476..166W,evans08}, with a similar value inferred
from the global star formation efficiencies in other galaxies (see,
e.g., \citealt{2007ARA&A..45..565M}).

One class of explanation for this low star formation efficiency is
that gas in the ISM is prevented from collapsing by, for example,
magnetic fields, cosmic rays, or by externally-driven turbulence
(Parker 1969; Sellwood \& Balbus 1999; Ostriker, Stone, \& Gammie
2001).  A second class of explanation is known by the name of
``feedback:'' the injection of energy and momentum into the ISM by
stellar processes so that star formation alters the ISM conditions and
limits the rate at which gas turns into stars. The form the feedback
takes is not currently agreed upon. Suggested mechanisms include
supernova heating, deposition of momentum by supernovae, heating by
photoionizing radiation from massive stars, deposition of momentum by
expanding bubbles of photoionized HII region gas, deposition of
momentum by the shocked winds from massive stars, and jets from
protostars (e.g., McKee \& Ostriker 1977; Silk 1997; Wada \& Norman
2001; Matzner 2002; Li \& Nakamura 2006; Cunningham et al.~2008).

In this paper, we study these feedback processes and assess the role
that they play in disrupting GMCs across a wide range of star-forming
galaxies.  In addition, we focus on a somewhat less well studied form
of feedback: deposition of momentum by the absorption and scattering
of starlight by dust grains \citep{odell, chiao72, elmegreen83,
  ferrara93, scoville01, scoville03, thompson05}.  Although the
magnetic fields in starburst galaxies can be large ($\sim$\,few mG for
Arp 220; Thompson et al.~2006, Robishaw et al.~2008), we neglect them
throughout this paper in order to focus on the competition between
various processes that contribute to disrupting GMCs.

It is important to distinguish between two arenas in which galactic
feedback likely operates: galactic disks in the large, and in the main
units of star formation, GMCs. While there is rather sharp debate in
the literature, we will assume that GMCs are at least marginally
gravitationally bound objects, and hence that they are unlikely to be
supported by feedback acting on the scale of galactic disks as a
whole.  As noted above, observations in our own and nearby galaxies
establish that only $\sim5\%$ of the gas in a GMC ends up in stars;
the rest of the gas is dispersed back into the ISM. The universality
of the Kennicutt law suggests that a similarly small fraction of the
GMCs in other, more distant classes of galaxies, is turned into
stars. Something is disrupting GMCs, but it is unlikely to be large
scale turbulence in the galaxy as a whole.  Instead, GMCs must be
disrupted by the stars that form in them.  A number of authors have
argued that galactic GMCs are disrupted by expanding HII regions
(e.g., \citealt{matzner02,krumholz06}); this mechanism cannot,
however, work in luminous starbursts \citep{matzner02}.  The fact that
these galaxies nonetheless have roughly similar star formation
efficiencies suggests that another disruption mechanism must be
competitive with expanding HII regions.

In this paper, we argue that the radiation pressure produced by the
largest few star clusters in a GMC, acting on dust grains in the gas,
is the primary mechanism by which GMCs are disrupted in more luminous
starbursts and massive GMCs (see also Scoville et al.~2001, Harper-Clark \& Murray
2009, and Pellegrini et al.~2007, 2009).
Protostellar jets also provide an important contribution, particularly
in the early stages of the evolution.  In spirals like the Milky Way,
both expanding HII regions and radiation pressure are comparably
important, depending on the size and mass of the cluster, and
supernovae also play an important role in the latest stages of
disruption.

\subsection{Is Feedback Really Necessary?}
\vspace{-0.04in}
\label{km}
A key thesis of this work is that stellar feedback is crucial for
understanding the low observed values of the star formation efficiency
in galaxies.  In contrast, \citet{krumholz} present an explanation of
the Kennicutt-Schmidt law (eq.~\ref{eqn: Kennicutt}) that does not
invoke an explicit form of feedback. Their argument is that turbulent
motions prevent the collapse of the bulk of the gas in a GMC (or in
other bound objects); only if the density is above a critical density,
which depends on the Mach number of the flow, do stars actually
form. The fraction of gas in a turbulent flow that lies above this
critical density is small, leading to the low observed star formation
efficiency per dynamical time.

We find this argument to be compelling, as far as it goes. As long as
turbulence is maintained, only a small fraction of gas will collapse
into stars per dynamical time. However, the assumption of a constant
level of turbulence is essential to the \citet{krumholz} argument.  A
key, and yet unanswered, question is thus {\it what maintains the
  turbulence?} If the turbulence in a GMC is not maintained, then the
GMC will contract, leading to an increase in the mean density and a
decrease in the dynamical time. Indeed, simulations find that
turbulence decays on $\sim$\,1 crossing time \citep{maclow,osg}, so
that a continued source of energy is needed to maintain the turbulent
support of the cloud.  It is possible, in principle, that
gravitational contraction of a GMC can sustain the turbulence,
maintaining approximate virial equilibrium and a slow contraction of
the cloud \citep{krumholz06}.  We argue, however, that an independent
internal source of turbulence, provided by stars, is crucial to
maintaining the slow rate of star formation.

As an example, we apply this argument to Arp 220. The interstellar
medium of Arp 220 has a turbulent Mach number ${\cal M}\approx100$.
The fraction of a GMC (or any bit of molecular gas) that is
sufficiently dense to be converted into stars in a free fall time is
then $\simeq0.013-0.05$ for GMC's with a virial parameter
$\alpha_{\rm vir}=0.1-1$ (see Figure 3 of \citealt{krumholz}).\footnote{A
  viral parameter $\alpha_{\rm vir} = 1$ corresponds to a cloud that is
  just gravitationally bound, while the smaller value of $\alpha_{\rm vir}
  = 0.1$ corresponds to a cloud that is approaching free fall
  conditions.} According to this argument, a GMC will convert half its
gas into stars in ten to forty free fall times, reasonably consistent
with equation (\ref{eqn: Kennicutt}) for any $\alpha_{\rm vir}$.  However,
this assumes that the cloud does not collapse and reduce its free fall
time.  In reality, if turbulence can only maintain $\alpha_{\rm vir} \sim
0.1$, a cloud is likely to collapse, leading to a rapid increase in
density and a decrease in the free-fall time.  If the star formation
rate per free-fall time remains roughly constant, then the actual star
formation rate will increase rapidly with time, and most of the gas in
the cloud will be turned into stars in roughly one initial
(large-scale) free-fall time.  Thus the model of \citet{krumholz} for
the low star formation efficiency in galaxies relies critically on
maintaining sufficient levels of turbulence so that $\alpha_{\rm vir} \sim
1$. On larger scales --- above the characteristic GMC size --- the
equivalent argument is that the galactic disk must have Toomre $Q \sim
1$, as we discuss below.

\vspace{-0.07in}
\subsection{This Paper}
\vspace{-0.03in}

The remainder of this paper is organized as follows. In \S
\ref{sec:AE}, we collect a number of relevant astrophysical results
used in our modeling. In \S \ref{sec:One D}, we describe a simple
one-dimensional model for the disruption of GMCs which includes the
effects of HII gas pressure, proto-stellar jets, radiation pressure,
gas pressure associated with shocked stellar winds, and wind shock
generated cosmic rays (many of the details of how we model these
forces are given in Appendix \ref{appendix:Forces}).  In section \S
\ref{sec:results} we present the results of our numerical modeling of
GMC disruption in star forming galaxies.  To explore the wide range of
conditions seen in galaxies across the Kennicutt-Schmidt law, we
consider models for GMCs in the Milky Way, M82, Arp 220, and the $z
\sim 2$ galaxy Q2346-BX 482. In \S\ref{sec:discussion}, we discuss the
implications of our results, the origin of turbulence in galaxies, and
explain physically why radiation pressure is the only source of
stellar feedback in principle capable of disrupting GMCs across the
huge dynamic range in ISM conditions from normal galaxies to the
densest starbursts.

\section{ASTROPHYSICAL ELEMENTS}
\label{sec:AE}
In this section we collect several pieces of observations and physics
that we believe are relevant to star formation in galactic disks and
GMCs. We order these items according to the amount of support they
enjoy, from substantial to slim.  The key conclusion below is that a
significant fraction of all stars are formed in compact ($\sim$ few pc
radius) massive star clusters, that in turn reside in GMCs.  Given the
importance of a few star clusters that are small compared to the GMC
as a whole, a one dimensional model for GMC disruption is a reasonable
first approximation; this is presented in \S \ref{sec:One D}.

\subsection{Marginally Stable Disks ($Q\approx1$)}
Quirk (1972) showed that normal galaxies have gas disks with
$Q\approx1$.  Kennicutt (1989) refined this to the statement that {\em
  within the star forming part of normal galactic disks}, $1/4\lesssim
Q\lesssim0.6$. At large radii he found $Q>1$ and a lack of star
formation. More recently, Leroy et al.~(2008) studied star
formation in detail in 23 nearby galaxies. They found that if they
accounted for only the gas surface density, as done above, their disks
were stable, with $Q\approx3-4$; using the total (gas plus stars) surface
density resulted in $Q\approx2$ with a slight variation in
radius. Unlike \citet{Kennicutt}, they find that star formation occurs
at large radii, beyond Kennicutt's $Q>1$ radius, albeit at reduced
rates.

These studies were restricted to normal galaxies, and employed a fixed
sound speed as the estimate for the random velocity. However, there is
evidence that disks with turbulent velocity $v_T>> c_s$ also satisfy
$Q\sim 1$ when $v_T$ is used in evaluating $Q$ (e.g.,
\citealt{thompson05} and our discussion of the starbursts
M82, Arp 220, and Q2346-BX 482 in \S\ref{sec:results}). 
Motivated by these observations and by theoretical considerations, we
will assume that all star forming galaxies have $Q\approx1$.

\subsection{The Toomre Mass and Giant Molecular Clouds}
We assume that galactic disks initially fragment on the disk scale
height $H\approx (v_T/v_c)r$. The fragments will form gravitationally
bound structures with a mass given by the Toomre mass, $M_T \simeq \pi
H^2 \Sigma_g$. Near the location of the sun, the gas surface density
$\Sigma_g\approx2\times10^{-3}\g\cm^{-2}$ and $H\sim300\pc$, giving
$M_T\approx2\times10^6M_\odot$.

This scenario is consistent with observations of GMCs in our
galaxy; in the Milky Way half the gas is in molecular form in giant
molecular clouds with a characteristic mass of order
$5\times10^5M_\odot$ \citep{1987ApJ...319..730S}, but with a rather
wide range of masses. The number of clouds $N(m)$ of mass $m$ is
given by
\be \label{eq:GMC} %$
{dN\over dm}=N_0\left({m_0\over m}\right)^{-\alpha_G},
\ee %$
with an exponent $\alpha_G\sim 1.8$ \citep{1987ApJ...319..730S} or
$1.6$ \citep{1997ApJ...476..166W}, so that most of the mass is in the
largest clouds. As a cautionary note, we note that \citet{engargiola}
find $\alpha_G=2.6\pm 0.3$ in M33, which suggests that lower mass
clouds contribute a significant fraction of the total mass. In the
Milky Way the largest GMCs have masses of order
$\sim3\times10^6M_\odot$ \citep{1987ApJ...319..730S}, roughly
consistent with the Toomre mass. In the Milky Way, and possibly in
other galaxies, molecular clouds are surrounded by atomic gas with a
similar or slightly smaller mass.

The clouds appear to be somewhat centrally concentrated. We will often
employ a Larson-law density distribution,
\be %$
\rho(r)\sim r^{-1},
\ee %$
where $r$ is the distance from the center of the GMC. We also explored
isothermal models $\rho(r)\sim 1/r^2$; we find that such clouds are
slightly easier to disrupt than the less centrally concentrated
Larson-law clouds in the optically thick limit.

\subsubsection{But are there molecular clouds in ULIRGs?}
GMCs are observed in the Milky Way, and in nearby star-forming galaxies
such as M82. We see clumps of gas in ``chain'' or ``clump'' star-forming
galaxies at $z=2$, such as Q2346-BX 482 discussed below. These have
been interpreted as self-gravitating, i.e., as GMCs (e.g.,
\citealt{genzel08}).  However, we do not know of any direct evidence
for Toomre mass self-gravitating objects in ULIRGs. There is some
evidence against such objects: since the clouds are self gravitating,
they will have a slightly higher velocity dispersion than that of the
disk out of which they form.  Increasing the velocity dispersion will
alter the inferred gas mass (see Downes \& Solomon 1998). Putting too
much gas in gravitationally bound objects will increase the apparent
gas mass, possibly making it larger than the dynamical mass.

One the other hand, ongoing star formation is clearly seen in Arp
220. Star formation probably requires densities exceeding
$10^6\cm^{-3}$ to proceed. Thus there is evidence that {\em some} gas
is gravitationally bound.  Moreover, there are numerous massive
compact star clusters observed in Arp 220 \citep{wilson}, indicating
that massive, bound, and relatively compact accumulations of gas
existed in the recent past.  Motivated by these considerations, we
will assume that Toomre-mass self-gravitating objects exist in all
star-forming galaxies, including ULIRGs.

\subsection{Gas clump and stellar cluster mass distributions}

Most of the gas in Milky Way GMCs is diffuse ($n\lesssim
3\times10^2\cm^{-3}$), but a fraction of order $10\%$ is in the form
of dense gas clumps, with sizes around $1\pc$ \citep{L2} and masses
from a few tens to a few thousand solar masses. The clumps have a mass
distribution similar to that for clouds (eq.~\ref{eq:GMC}), with an
exponent $\alpha_{c}\approx1.7$ \citep{L2}. 

In both the Milky Way \citep{EE97,vdBL84} and in nearby galaxies
\citep{1997ApJ...476..144M,1989ApJ...337..761K}, the number of stellar
clusters of mass $m_*$ is given by
\be \label{eq: cluster dndm}%$
{dN_{*\rm cl}\over dm_{*\rm cl}}=N_{*\rm cl,0}\left({m_{*\rm cl,0}\over m_{*\rm cl}}\right)^{\alpha_{\rm cl}}
\ee %$
with $\alpha_{\rm cl}\approx1.8$.
In other words, most stars form in massive clusters; in the Milky Way,
at least, these clusters are made from gas in
massive gas clumps, inside of massive GMCs.

\subsection{The Sizes of Star Clusters}
\label{sec:size}

Star clusters are observed to have sizes ranging from
$\rcl\approx0.1\pc$ (for $\mcl\approx10M_\odot$) to $\rcl\approx10\pc$
($\mcl\approx10^8M_\odot$). There are hints that clusters with masses
$\mcl\lesssim10^4M_\odot$ have a mass-radius relation of the form
\be   \label{eq: cluster size small}%$
\rcl\approx 2\left({\mcl\over m_0}\right)^\beta\pc
\ee   %$
with $m_0=10^4$\,M$_\odot$ and $\beta\approx0.4$ \citep{L2}, but it is
entirely possible that this is a selection effect.  Intermediate mass
clusters, those with
$10^4M_\odot\lesssim\mcl\lesssim3\times10^6M_\odot$ have $\rcl \simeq
2$ pc independent of mass ($\beta=0$), albeit with substantial
scatter. This characteristic size is seen for young ($\lesssim
30\Myr$) and old ($\gtrsim 30\Myr$) clusters in M51 with masses in the
range $10^3-10^6M_\odot$ \citep{Scheepmaker}, for super star clusters
in M82 with $\mcl=10^5-4\times10^6M_\odot$
\citep{McCrady03,McCrady07}, and in globular clusters with masses
$\sim10^5-10^6M_\odot$ \citep{Harris}.  Finally, high-mass clusters
with $\mcl\gtrsim10^6M_\odot$ have $\beta=0.6$ and
$m_0=10^6$\,M$_\odot$ in equation (\ref{eq: cluster size small})
\citep{walcher05,evstigneeva07,barmby07,Rejkuba,murray09}.  When using
the radii of stellar clusters in our GMC models, we will be guided by
these observed mass-radius relations.

\begin{deluxetable}{lcccccc}
%\rotate 
\tablewidth{0pt} \tablecaption{OBSERVED GALAXY PARAMETERS}
\tablehead{
\colhead{Galaxy}    & \colhead{$R_d$}   & $t_{\rm dyn}$ & \colhead{$\Sigma$}   & \colhead{$v_T$}   & \colhead{$Z/Z_\odot$} & \colhead{$\dot M_*$ (obs)} \\
\colhead{\phantom} & \colhead{$\kpc$} & \colhead{$\yr$} & \colhead{${\rm g\,\,cm^{-2}}$} &  \colhead{$\kms$} & \colhead{\phantom}  &\colhead{$M_\odot\, \yr^{-1}$}} \\

\startdata
Milky Way   &  $8.0$  & $3.6\times10^7$  &  $2\times10^{-3}$ & $6$    & $1$   & $2.0$ \\
M82         &  $0.35$ & $3.0\times10^6$  &  $0.1$            & $15$   & $1.5$ & $4$ \\
BX482       &  $7.0$  & $2.9\times10^7$  &  $4\times10^{-2}$ & $53$   & $1$   & $140$  \\
Arp 220     &  $0.1$  & $3.3\times10^5$  &  $7$              & $61$   & $3$   & $120$ \\

\enddata

%\tablenotetext{a}{Any notes go here} 
\tablecomments{Observed galaxy properties. Column one gives the name
  of the model. The next five columns give model input parameters: the
  disk radius (col. 2), dynamical time $R_d/v_c$ (col. 3), gas surface
  density (col. 4), turbulent velocity $v_T$ (col. 5; recall that
  $H=[v_T/v_c] R_d$), and metallicity in solar units (col. 6). The
  metallicity is not that well-constrained in Arp 220.  Column 7 gives
  the observed star formation rate for the galaxies. }

\end{deluxetable}

\begin{deluxetable}{lcccccc}
%\rotate 
\tablewidth{0pt} \tablecaption{GMC AND STAR CLUSTER PROPERTIES}
\tablehead{
\colhead{Galaxy}    & \colhead{$\Rg$} & \colhead{$\Mg$}   & \colhead{$\rcl$} & \colhead{$M_*$} & \colhead{$\eg$}     & \colhead{$v_T$} \\
\colhead{\phantom} & \colhead{$\pc$} & \colhead{$M_\odot$} & \colhead{$\pc$} & \colhead{$M_\odot$} & \colhead{\phantom} & \colhead{$\kms$}} \\

\startdata
Milky Way          &  $100$         & $3\times10^6$      & $2$              & $10^5$            &   $0.03$            &  $5$  \\
M82                &  $23$          & $3\times10^6$      & $1.5$            & $7\times10^5$     &   $0.24$            &  $10$  \\
BX482              &  $925$         & $10^9$             & $13$             & $2.7\times10^8$   &   $0.27$            &  $50$  \\
Arp 220            &  $5$           & $4\times10^7$      & $3.5$            & $1.4\times10^7$   &   $0.38$            &  $50$  \\

\enddata

%\tablenotetext{a}{Any notes go here} 
\tablecomments{Columns 2-5 give the assumed GMC and star cluster
  properties: the radius of the GMC $\Rg$ (col. 2), the mass of the
  GMC $\Mg$ (col. 3), the star cluster radius $\rcl$ (col. 4), and the
  stellar mass of the star cluster $M_*$ (col. 5). Columns 6 and 7
  give the predictions of our model for the star formation efficiency
  in the GMC $\eg$ and the shell velocity when the GMC is disrupted,
  which we also interpret as the turbulent velocity $v_T$ induced in
  the ISM of the Galaxy.}

\end{deluxetable}

\section{A MODEL OF CLUSTER \& GMC DISRUPTION}
  \label{sec:One D}

  We start by specifying the properties of the disk in which the GMC
  lives; the effective disk radius $R_d$, the circular velocity $v_c$,
  the disk scale height $H_d$, the disk gas mass $M_d$ and metallicity
  $Z/Z_\odot$ (relative to solar). We choose the values of these
  parameters to match those of four galaxies to which we compare our
  models: the Milky Way, M82, Q2346-BX 482, and Arp 220. In that sense
  these are not free parameters.  Table 1 summarizes the observed
  input parameters for the galactic disks in the systems we model,
  while Table 2 gives the inferred or assumed input parameters related
  to the GMC and its central star cluster: the mass $\Mg$ and radius
  $\Rg$ of the GMCs and the stellar mass and radius of the star
  clusters. Table 2 also lists the GMC's star formation efficiency
  $\eg$ and the shell velocity when the GMC is disrupted -- we
  interpret the latter as the turbulent velocity induced in the ISM,
  $v_T$.

  We employ a one-dimensional model for the GMC, with the free
  parameter $\phi_G\equiv H_d/\Rg$ defining the size of a GMC with
  respect to the disk in which it resides. In the Milky Way this ratio
  is about $\sim2-5$ for the most massive clouds; for specific cases
  like G298.4-0.3, which has $\Rg\approx100\pc$, we use the observed
  ratio. In models for other galaxies we fix $\phi_G=4$. The mass of
  the GMC is taken to be the Toomre mass, with a Larson-like
  ($\rho\sim1/r$) internal density profile.

  For the purposes of our simplified modeling, the stars are assumed
  to lie in a single massive cluster of total mass (gas plus stars)
  $M_{\rm cl}$, which forms a mass of stars $M_*=\ecl M_{\rm cl}$ with
  luminosity $L$, surrounded by the remnants of the gas out of which
  it formed, with mass $M_g=(1-\ecl)M_{\rm cl}$.  We use a
  \citet{muench02} stellar initial mass function (IMF) to relate the
  cluster luminosity to its mass.  The quantity $\ecl$ characterizes
  the efficiency with which cluster gas is turned into stars. In our
  galaxy $\ecl\approx0.3-0.5$ \citep{L2}.  In our models, we fix $\ecl
  \simeq 0.5$ and adjust the cluster mass $\mcl$ (or equivalently, the
  cluster stellar mass $M_*$) to find under what conditions the
  central star cluster can disrupt its host GMC.  Physically, we
  expect that star formation will self-adjust to a form a cluster of
  approximately this mass.  For the Milky Way and M82 the star cluster
  masses we infer by this method are comparable to those observed.

We model the impact of the central star cluster on the surrounding GMC
using the thin shell approximation.  As the star cluster evolves, 
driving winds, jets, and radiation into the overlying shell of gas, 
we calculate the shell's dynamics as it sweeps up mass and disrupts the GMC.  
The relation between shell radius
$r$, shell velocity $v(r)$, shell mass $M(r)$, and shell momentum
$P_{\rm sh}(r)$ is given by
\be %
{dr\over dt}= \frac{P_{\rm sh}}{M(r)},
\ee %
where $M(r)$ is the mass of the shell, which increases as the shell
radius $r$ increases, and $P_{\rm sh}=M(r)v(r)$ is the momentum of the
shell.  For a Larson-like density profile of the GMC, $\rho \propto
r^{-1}$, the mass of the shell is given by
\be \label{eq:mass} %
M(r)=  (1-\ecl) M_{\rm cl}+M_{\rm GMC}
\left[
\left({r\over\Rg}\right)^2
-\left({r_{\rm cl}\over\Rg}\right)^2
\right]
\ee %$
for $r<R_{\rm GMC}$ and by
\be \label{eq:mass_d}%$
M(r)=(1-\ecl) M_{\rm cl}+M_{\rm GMC} + {4\pi\over3} (r^3-R_{\rm GMC}^3)\rho_{\rm disk} 
\ee %
for $R_{\rm GMC}<r<H$. We have experimented with both isothermal
($\Mg(r)\sim r$) and Larson ($\Mg(r)\sim r^{2}$) GMCs and find
qualitatively similar results.

The momentum equation for the shell is
\be \label{eq: momentum} %
{dP_{\rm sh}\over dt}=-F_{\rm grav}-F_{\rm ram}-F_{\rm turb}+F_{\rm HII}+F_{\rm rad}+F_{\rm jet}+F_{\rm hot}+F_{\rm cr}
\ee %
The inward forces on the right-hand side of equation (\ref{eq:
  momentum}) are the self gravity $F_{\rm grav}$ of the swept up shell and
the mutual gravity of the stars in the largest cluster and the shell,
the force $F_{\rm turb}$ exerted by the turbulent motions of the gas in
the GMC on the shell, and the force $F_{\rm ram}$ produced by ram pressure
as the shell sweeps up the material of the GMC or the surrounding gas
disk.  The outward forces on the right-hand side of equation (\ref{eq:
  momentum}) acting to disrupt the GMC include the force $F_{\rm jet}$
associated with momentum deposited by jets from star formation, the
force $F_{\rm HII}$ due to the thermal pressure from ionized gas (HII
regions) and from shocked stellar winds, and the force $F_{\rm rad}$
associated with radiation pressure. In some models we include the
effects of hot gas $F_{\rm hot}$ and cosmic rays $F_{\rm cr}$ associated with
shocked stellar wind.  A more detailed description of how we implement
these forces is given in Appendix \ref{appendix:Forces}.

We assume that stellar wind energized hot gas ($\sim10^7$K) does not
play a dynamical role in galaxies like the Milky Way; we show
explicitly in this paper that even in the most optimistic cases such
winds are not important for GMC disruption in ULIRGs. Much of the
literature on bubbles around massive stars and massive star clusters
assumes that shocked stellar winds dominate the dynamics (e.g.,
\citealt{1975ApJ...200L.107C,1977ApJ...218..377W,chu90}). However,
observations of HII regions in the Milky Way suggest that the pressure
in such hot gas is equal to that of the associated HII ($10^4$K) gas
\citep{DM87,McKee84,HCM}. The most likely interpretation of these
results is that neither hot gas nor cosmic rays are confined inside
bubbles in the Milky Way or the LMC, but rather escape
\citep{HCM}. Accordingly, we neglect the last two terms on the right
hand side of equation (\ref{eq: momentum}) in our Milky Way models and
in our M82 models; calculations which include these pressures predict
bubble sizes that are far too large compared to observations in the
Milky Way.  Accordingly, with the exception of \S \ref{sec:arp220}, we
ignore the pressure associated with shocked stellar winds. We do
include stellar-wind and cosmic-ray pressure in our Arp 220 models,
but there they make little difference.

In \S\ref{sec:intro}, we made a distinction between feedback in the
disk as a whole, and feedback in GMCs. This distinction is important
for a number of reasons.  In particular, we believe that supernova
(SN) explosions largely contribute to the former, but not the latter.
In $z \sim 0$ ULIRGS, where the bulk of the star formation takes place
in $\sim100\pc$ disks, the dynamical time is $t_{\rm
  dyn}=R/v_c\approx5\times10^5\yr$, much less than the main sequence
lifetime of even the most massive stars. Hence a gravitationally bound
object in a ULIRG cannot be disrupted on a dynamical time by SNe
resulting from stars formed in that object.  In the Milky Way, even
though the dynamical time is longer than the main sequence lifetime of
massive stars, it is clear that GMCs are in the process of being
disrupted well before SNe occur \citep{MR}.  Although SNe may be
important during late stages of GMC disruption, and for stirring up
the galactic disk as a whole, they cannot be the main agent that
disrupts GMCs, either in Milky Way-like galaxies or in ULIRGs.  For
this reason, we do not include the force due to SNe in equation
(\ref{eq: momentum}).

In the densest starbursts, which have mean gas densities $\sim
10^{3-4}$ cm$^{-3}$, SNe rapidly lose the majority of their energy to
radiative losses (e.g., Thornton et al. 1998).  Under these
conditions, the primary role of SNe is to stir up the bulk of the ISM
via the momentum they supply, rather than to heat up the gas and/or
create a hot phase of the ISM in pressure equilibrium with the rest of
the mass \citep{thompson05}.  To see that the latter is untenable, we
estimate the density $n_h$ of hot gas required for a virialized ISM at
$\sim 10^7 T_7$ K to be in pressure equilibrium with the bulk of the
gas, i.e., for $p_h = n_h k T \simeq \pi G \Sigma_g^2$.  Given this
density, we find that there is a critical surface density $\Sigma_c$
above which thermal X-ray emission from the hot gas would exceed the
observed correlation between X-ray and FIR emission ($L_X \simeq
10^{-4} L_{\rm FIR}$; \citealt{ranelli03}): $\Sigma_c \simeq 0.04
(H_d/100 \, {\rm pc})^{-2/5} T_7^{3/5}$.  Galaxies with $\Sigma_{\rm
  d} \gtrsim \Sigma_c$, which includes the majority of luminous
star-forming galaxies, cannot have a dynamically important hot ISM.
Instead, the dominant role of SNe is to stir up the dense gas via the
momentum imparted in the snowplow phase.  This may even be true in
galaxies with $\Sigma_c \lesssim \Sigma_d$, because the hot gas can
vent via galactic winds or fountains.  For example, in the Milky Way,
which has $\Sigma_d \simeq 2.5 \times 10^{-3}$ g cm$^{-2}$ $\ll
\Sigma_c$, the hot ISM is believed to contribute only $\sim 10 \%$ of
the total pressure \citep{boulares}.  For this reason, we will not
consider the pressure due to the hot ISM in this paper, although the
momentum supplied by SNe is important in the late stages of star
cluster and GMC evolution.

Having summarized the basic elements of our model, we now describe its
application to galaxies ranging from the Milky Way to the most
luminous starbursts (\S \ref{sec:results}).  We then discuss the
implications of these results (\S \ref{sec:discussion}).

%--------------------------------------------------------------------
\begin{figure*}
\centerline{\psfig{file=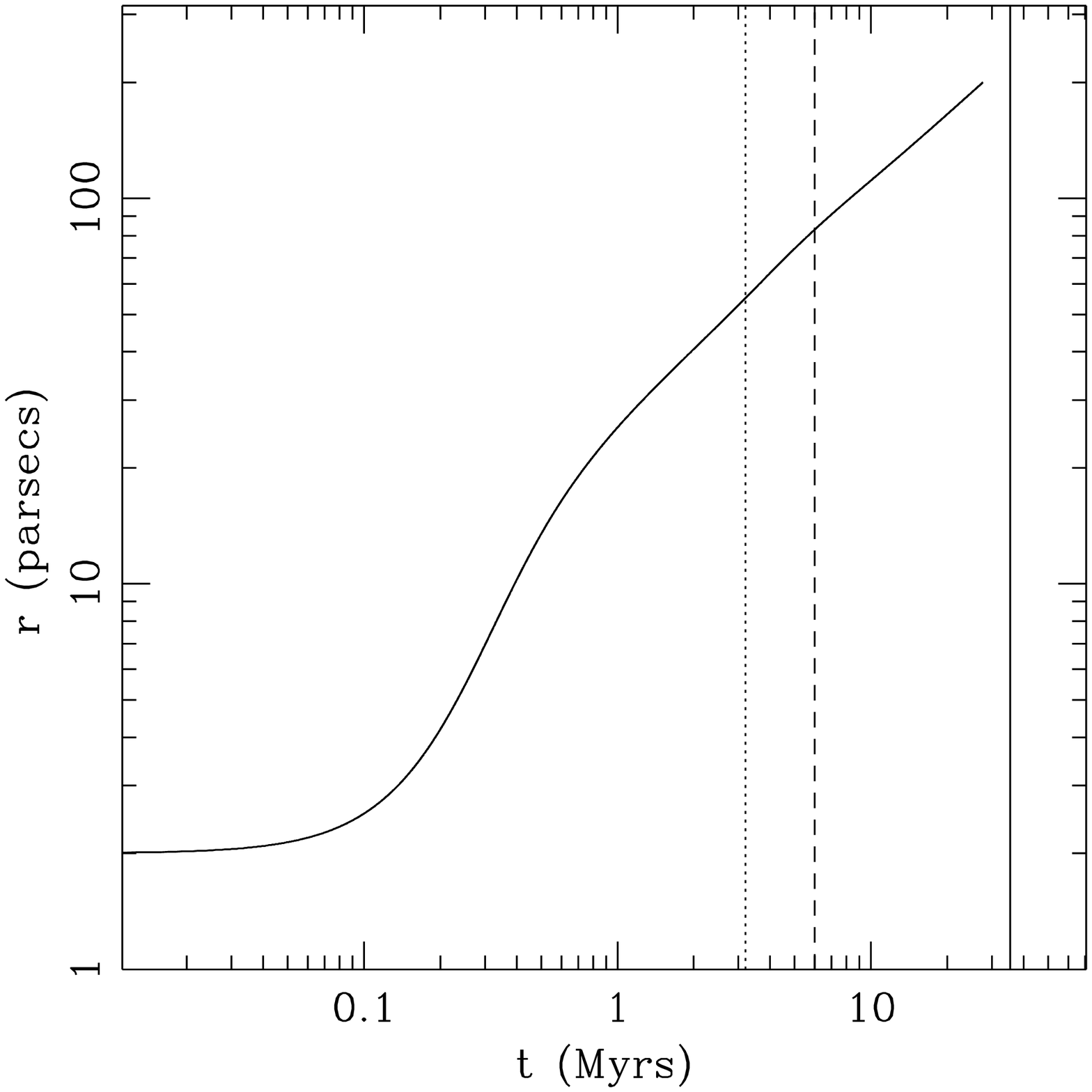,width=7.5cm}\psfig{file=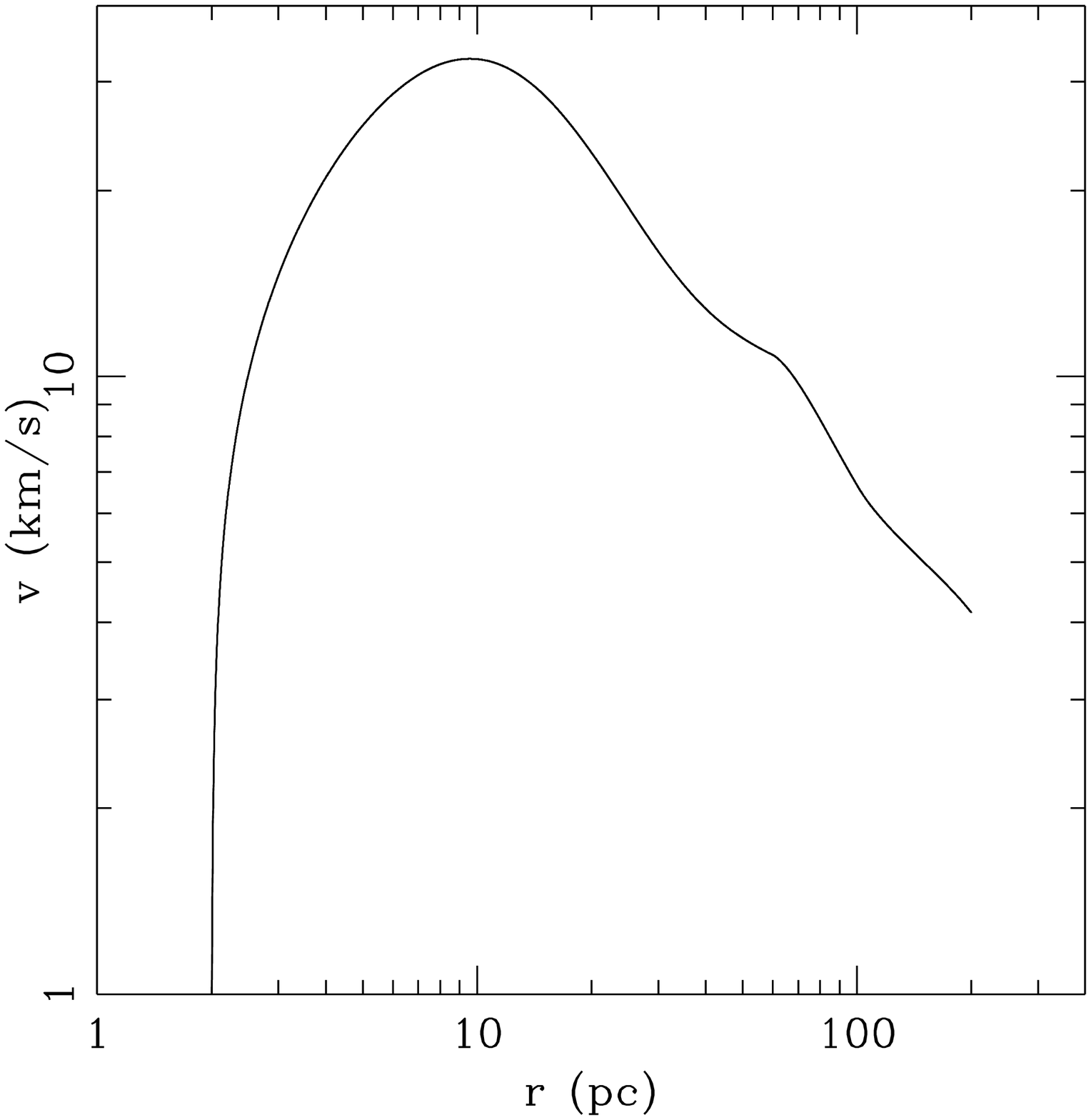,width=7.5cm}}
\centerline{\psfig{file=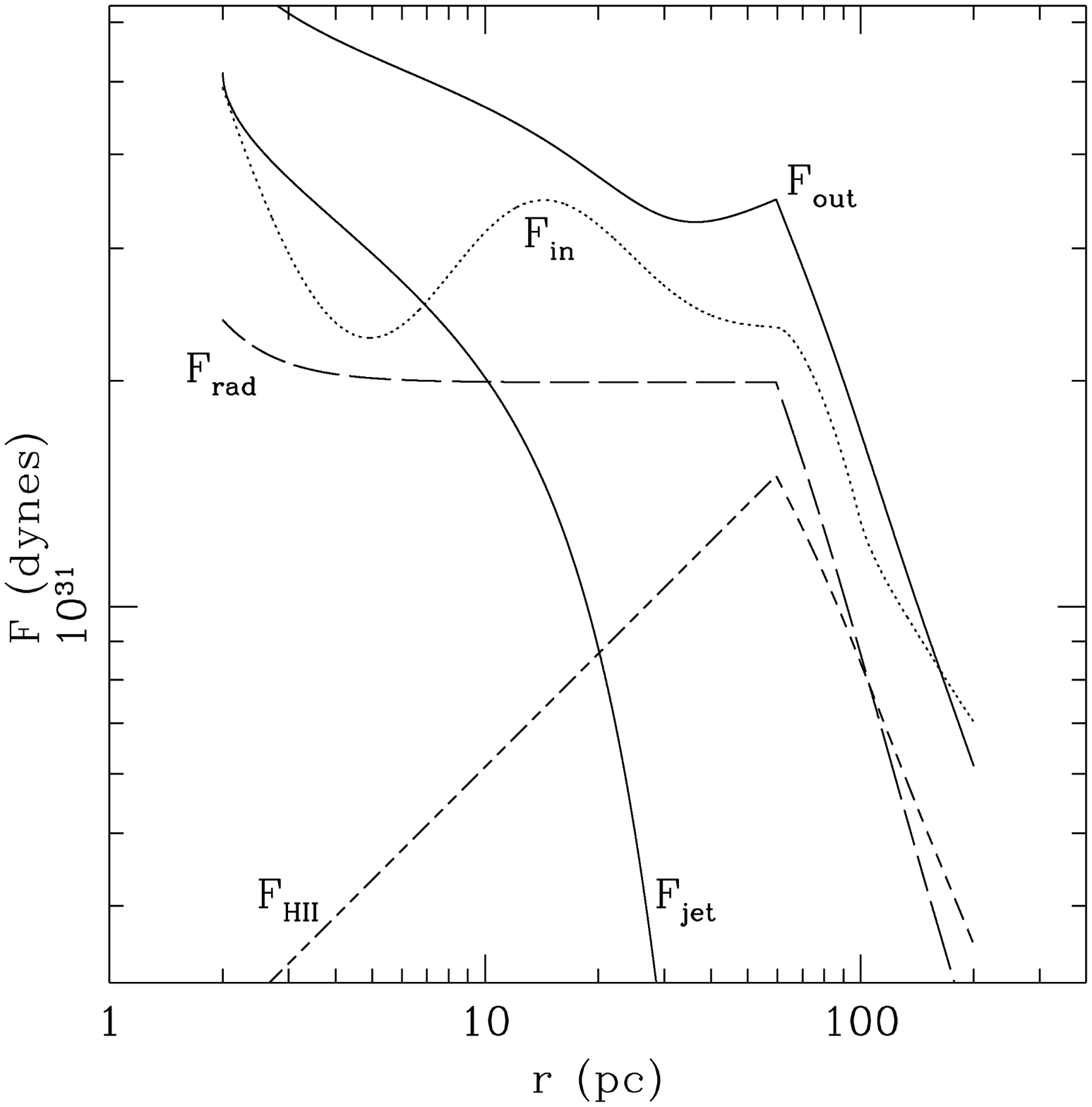,width=7.5cm}\psfig{file=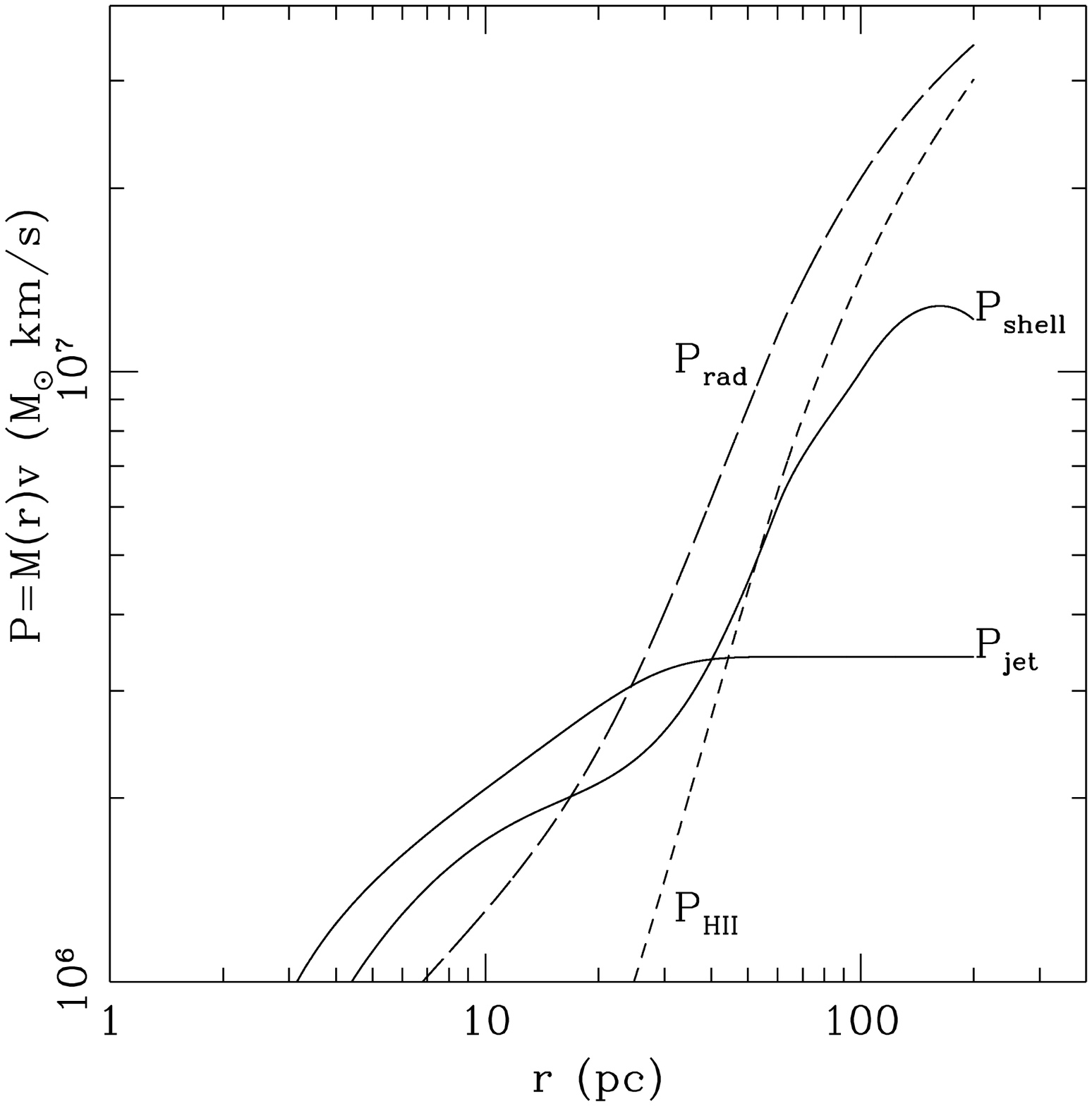,width=7.5cm}}
\caption{Shell radius as a function of time and velocity, forces, and
  shell momentum as a function of radius in our model for G298.4-0.3
  in the Milky Way.  The shell is sheared apart when it reaches the
  Hill radius ($\sim200\pc$), where we end our integration.  {\it
    Upper left:} The dotted, dashed, and solid lines mark when the
  first cluster SNe explode, the central cluster luminosity drops to
  $1/3$ of its initial value, and $t$ reaches the Milky Way dynamical
  timescale, respectively.  {\it Upper right:} The velocity of the
  swept-up shell in G298.4-0.3 as a function of shell radius.  Note
  that the asymptotic velocity is comparable to the turbulent velocity
  of the Milky Way disk.  {\it Lower left:} The upper-most solid line
  is the total outward force, consisting of the momentum supplied by
  protostellar jets (solid line; $F_{\rm jet}$), by HII gas pressure
  (short dashed line; $F_{\rm HII}$) and radiation pressure on dust
  grains (long dash line; $F_{\rm rad}$). The dotted line is the total
  inward force ($F_{\rm in}$), dominated by the self-gravity of the
  shell.  {\it Lower right:} Momentum of the swept-up shell in
  G298.4-0.3 (solid line), together with the momentum deposited by
  radiation (long dashed line), gas pressure (short dashed line), and
  protostellar jets (solid line, labeled $P_{\rm jet}$). The bulk of
  the momentum is supplied by radiation and gas pressure, but the
  early contribution of the protostellar jets is important in the
  disruption of the natal cluster gas.}
\label{fig:mw}
\end{figure*}
%-----------------------------------------------------------------------

\section{RESULTS}
\label{sec:results}

\subsection{The Milky Way: G298.4-0.3}
\label{sec:MW}

Recent work has revealed that the Milky Way harbors a number of very
young massive star clusters, with masses ranging up to $10^5M_\odot$
\citep{figer99,brandner,figer06,MR}.  We examine the case of G298.4-0.3
in the Carina arm. \citet{MR} show that the free-free flux emerging
from this region implies a total ionizing flux
$Q=7.7\times10^{51}\s^{-1}$, and suggest that $\sim60\%$ of this comes
from a single massive cluster residing in the prominent $\sim50\pc$
bubble revealed by Spitzer GLIMPSE images. Most of the remaining flux
comes from two clusters associated with the two giant HII regions
G298.2-0.3 and G298.9-0.4; both sources appear to lie in the rim of
the bubble seen in the GLIMPSE images. \citet{grabelsky} find two
massive GMCs in this region, numbers 24 and 26, both with
$\Mg\approx3\times10^6M_\odot$ and $\Rg\approx100\pc$. Their radial
velocities are $22\kms$ and $24\kms$, in good agreement with the range
of radio recombination line, i.e., HII region, radial velocities in
this direction, which range from $+16\kms$ to $30.3\kms$, with a mean
$\sim+23\kms$.

Accordingly, our model for G298.4-0.3 consists of a GMC with
$\Rg=100\pc$ and $\Mg=3\times10^6M_\odot$. In the spirit of our
simplified one-dimensional modeling, we lump all of the star clusters
together into a central star cluster with $L\approx7\times10^7L_\odot$
and initial cluster radius $\rcl=1.5\pc$.  Half the Galactic star
formation takes place in 17 star clusters, with a minimum
$Q=3\times10^{51}$, so G298.4-0.3 is representative of star forming
clusters in the Milky Way.

The top two panels of Figure \ref{fig:mw} show the radius of the shell
surrounding the central cluster as a function of time and the shell
velocity as a function of radius. The shell starts at our putative
initial radius for the cluster of $\sim1.5\pc$, and reaches
$r\sim80\pc$ at about 6.5\,Myr (dashed line), at which point the star
cluster luminosity has dropped by a factor of $3$.  The most massive
stars begin to explode after about $3.6$ Myr (vertical dotted line),
while the last O stars explode after about $1.3\times10^7\yr$.  The
solid vertical line marks the dynamical time $R/v_c$ for the Milky Way
at $R=8\kpc$. Note that the dynamical time for the GMC is somewhat
shorter, $\sim6$\,Myr.

The lower left panel of Figure \ref{fig:mw} shows the forces as a
function of radius in this model; note that the radiation and HII gas
pressures drop after $6.5$ Myrs, when the bubble has $r \sim 100$ pc,
but the bubble continues to expand at the same rate. The evolution
after $6.5$ Myrs, i.e., radii larger than $\sim50\pc$, may
underestimate the rate of expansion somewhat; the bubble may expand
slightly more rapidly after several Myrs due to energy input by
SNe. We say the rate of expansion {\em may} be underestimated since
the hot gas from the SNe is likely to escape the bubble as easily as
the hot gas from shocked winds apparently does. In addition to SNe,
other unmodeled effects also become important at late times and large
radii. For example, because the inner radius of the shell exceeds the
outer radius of the initial GMC, the surface density of the gas
decreases to $A_V\approx1$, so that ionizing photons from the Galactic
radiation field can penetrate and ionize the shell.

We halt the integration in our model when the radius of the expanding
bubble exceeds the Hill radius $r_{\rm Hill}$ of the GMC, i.e., when the
tidal shear from the Galaxy exceeds the self-gravity of the GMC:
$r_{\rm Hill}\approx(\Mg/2M(r))^{1/3}a$, where $M(r)=v_c^2a/G$ is the
enclosed dynamical mass of the galaxy at the galactocentric radius $a$
of the GMC. After this time the remaining molecular gas will be
dispersed (although not necessarily converted to atomic gas).

Figure \ref{fig:mw} shows that the central cluster in
G298.4-0.3 should disrupt its natal GMC. What force is responsible for
this disruption?  At the current radius, $\sim 55\pc$, the
radiation force and the gas pressure force are within a factor of two
of each other, and will remain so until most of the O stars explode;
the force from protostellar jets is substantially smaller. However, at
early times, the jet force was as much as a factor of two larger than
the radiation pressure force, and the gas pressure force was
negligible.

Finally, the lower right panel of
Figure \ref{fig:mw} plots the momentum of the shell as a
function of time. At the current radius of the bubble in G298.4-0.3,
$r\sim55\pc$, the radiation has deposited about twice the momentum
supplied by the HII gas pressure. The stellar jets are not active at
this time, but over the time they were active (corresponding to radii
below $\sim15\pc$) they deposited a momentum comparable to that of the
radiation pressure (at those early times). In these models, a combination
of proto-stellar jets and radiation pressure disrupts the natal
cluster, while a combination of gas and radiation pressure disrupts
the GMC.  The shell velocity at late times is of the order of the
turbulent velocity seen in the ISM of the Galaxy (upper right panel), 
demonstrating that, even in the absence of supernovae, massive
star formation can generate turbulent motions on large ($50\pc$ or
larger) scales comparable to those observed.

\subsection{The Starburst M82}

M82 is one of the nearest ($D=3.6$ Mpc; \citealt{Freedman94})
starburst galaxies, with an infrared luminosity
$L_{\rm IR}=5.8\times10^{10}L_\odot$ \citep{Sanders03}. The galaxy is
small compared to the Milky Way, with a circular velocity
$v_c\approx110\kms$ \citep{YoungScoville}, and a CO inferred gas mass
$2\times10^8M_\odot$ inside $r=350\pc$ \citep{Weiss01} (adjusted to
our assumed distance), yielding a gas surface density
$\Sigma_g\approx0.1\g\cm^{-2}$ and a gas fraction $f_g\approx0.2$. The
metallicity is $1.2-2.0$ times solar \citep{Smith
  Westmoquett2006MNRAS.370..513S}.

The radius and mass of the most massive star clusters in M82 are well
established; there are about $200$ clusters with $M>10^4M_\odot$
\citep{melo} and about $\sim 20$ well studied super star clusters
($\mcl>10^5M_\odot$).  With one arcsecond corresponding to a spatial
scale of $17.5\pc$, a number of super star clusters are resolved by
HST \citep{SmithGallagher,McCrady03,McCrady07}. Typical half light
projected radii for these massive objects are $\sim0.08''$ or
$1.4\pc$. \citet{McCrady03} list 20 such clusters. The total mass of
the 15 clusters for which they measure viral masses is $\sim
1.4\times10^7M_\odot$. Their largest cluster, `L', is a monster, with
a mass of $4\times10^6M_\odot$ and a half-light radius of $1.5\pc$;
more typical masses are $\sim5\times10^5M_\odot$. A rough fit of the
form (\ref{eq: cluster dndm}) gives $\alpha_{\rm cl}\approx1.9$
\citep{McCrady07}.

The masses of the GMCs in M82 are also known; the distribution is well
fitted by equation (\ref{eq:GMC}), with $\alpha_G\approx1.5\pm0.1$,
and a maximum mass of $\Mg\approx3\times10^6M_\odot$ \citep{keto}. The
Toomre mass is $\approx7\times10^6M_\odot$. Both are comparable to the
mass of the two largest super star clusters given by
\citet{McCrady07}. Either there were more massive GMCs in M82 in the
past, or $\eg\approx1$ for the GMCs out of which these two clusters
formed.

\begin{figure*}
\centerline{\psfig{file=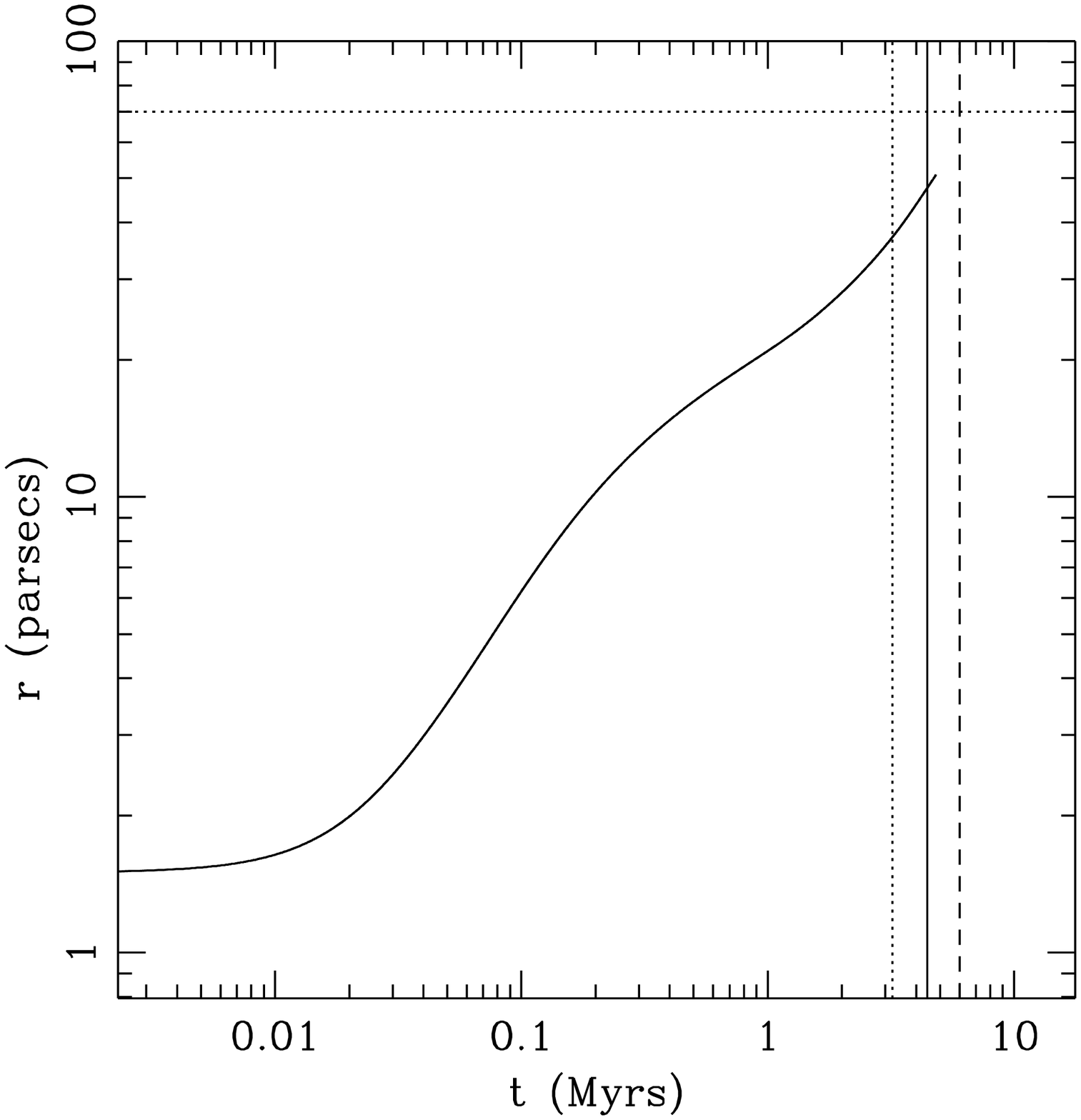,width=7.5cm}\psfig{file=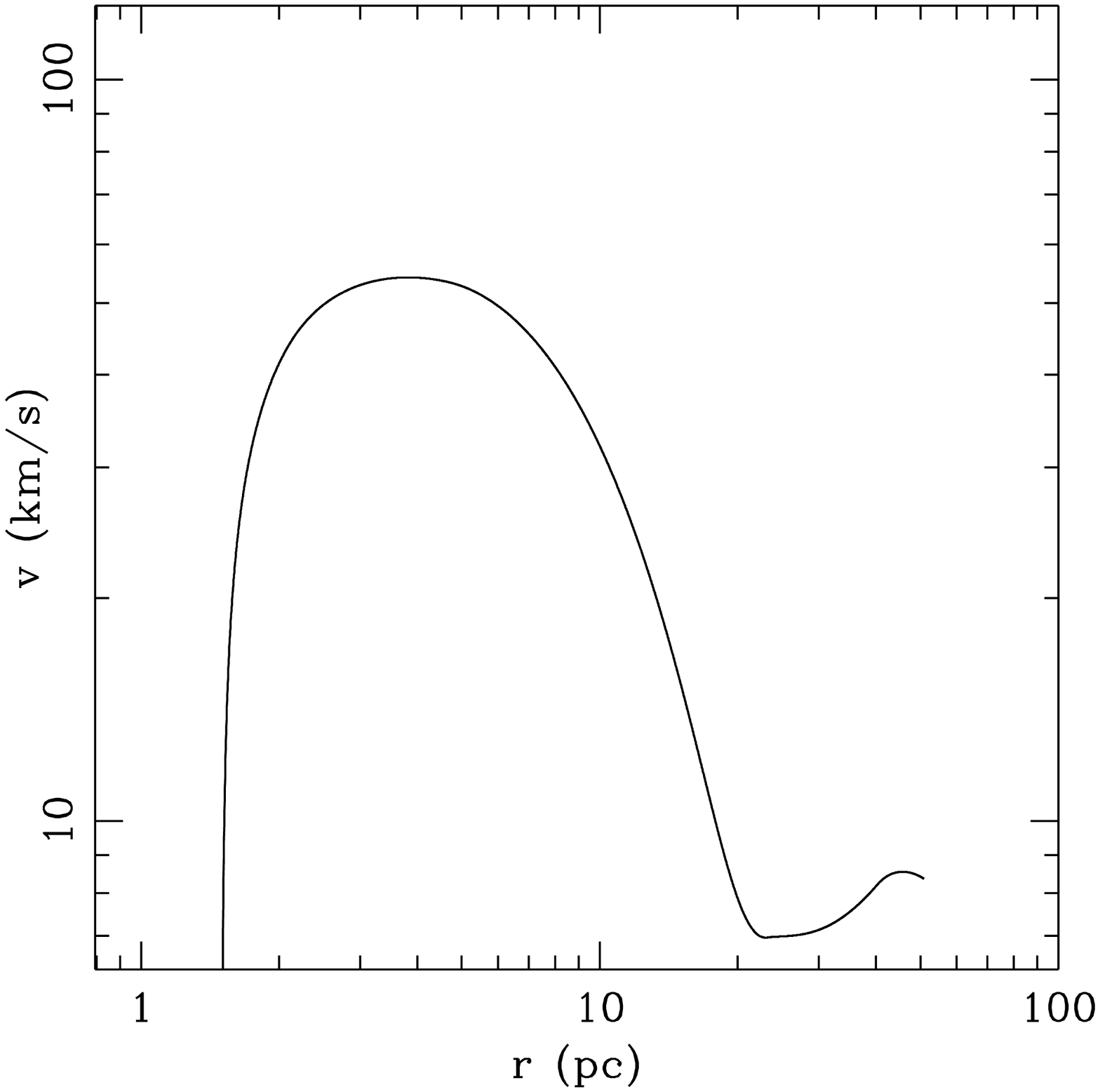,width=7.5cm}}
\centerline{\psfig{file=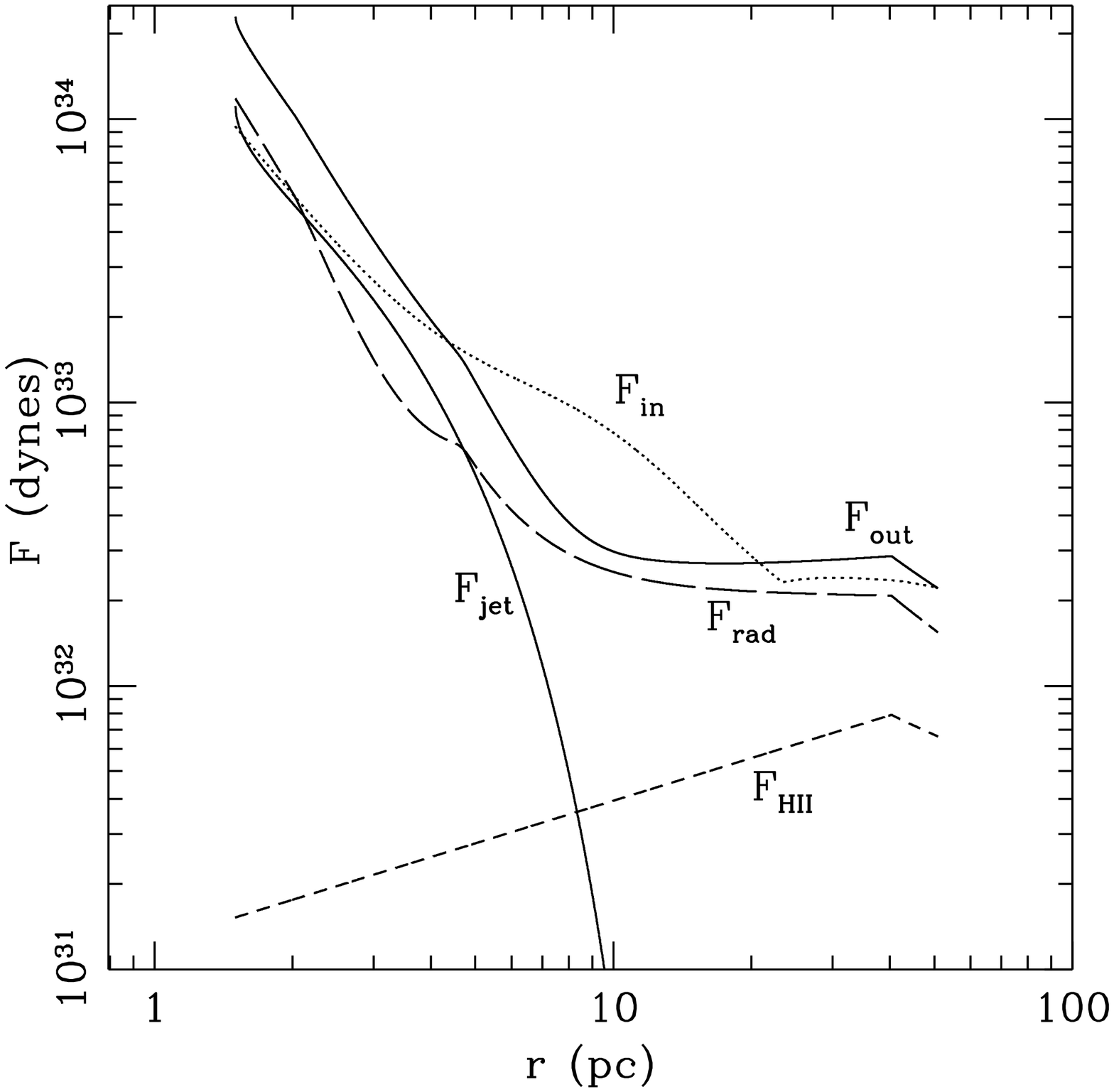,width=7.5cm}\psfig{file=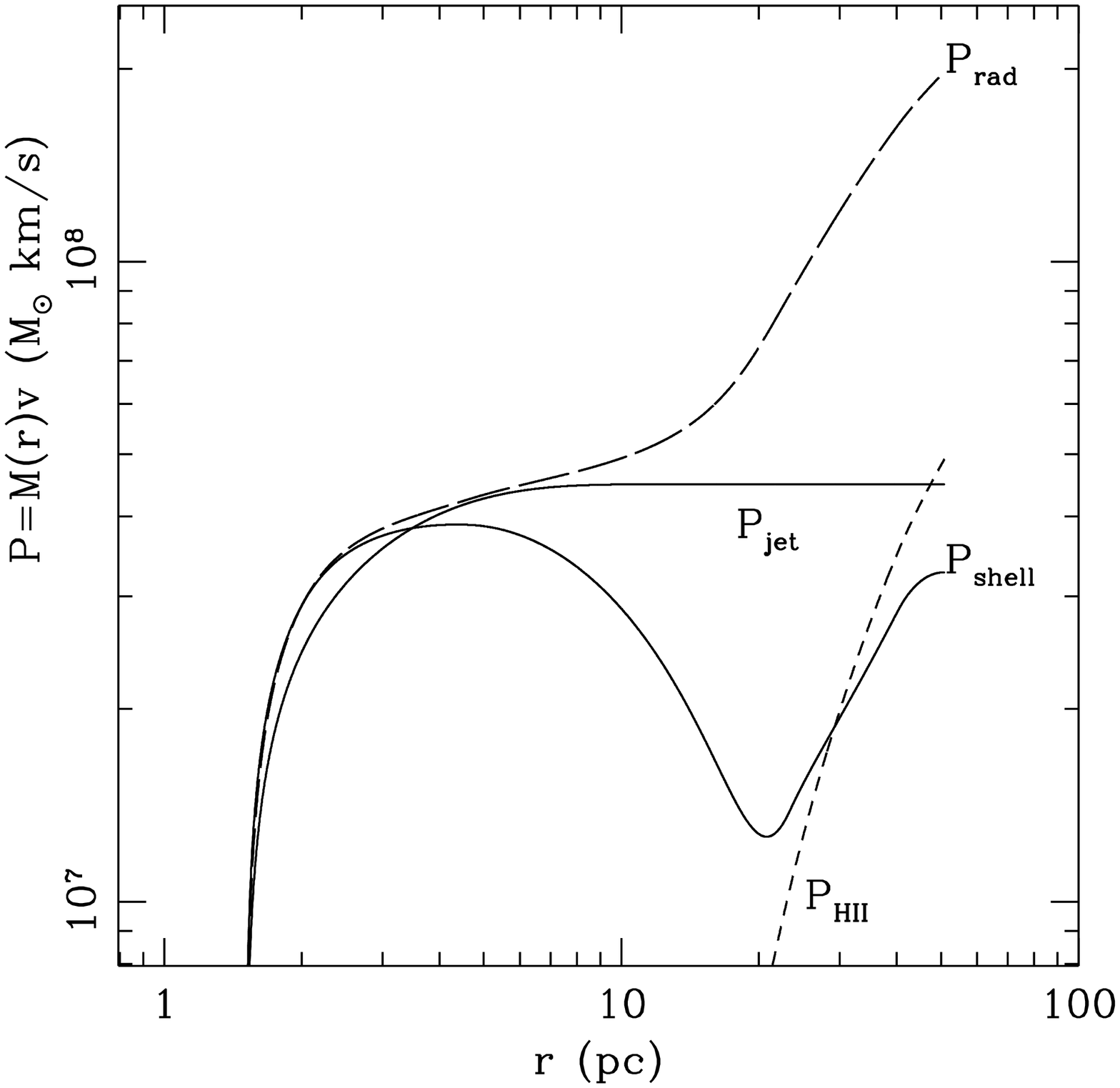,width=7.5cm}}
\caption{Shell radius as a function of time and velocity, forces, and
  shell momentum as a function of radius in our model for the
  disruption of a GMC by a massive star cluster in the starburst M82.
  {\it Upper left:} Bubble radius as a function of time. The dashed
  line is when the star cluster luminosity drops to 1/3 of its initial
  value, while the dotted vertical line marks the lifetime of a
  $120M_\odot$ star. The shell is disrupted by ambient differential
  rotation and tidal forces in the galaxy at the Hill radius (solid
  line); this occurs roughly when the first supernovae explode.  {\it
    Upper right:} The velocity of the swept-up shell in the M82 model
  as a function of shell radius.  {\it Lower left:} The forces in our
  M82 model, plotted against the radius of the swept up shell. Line
  styles are the same as in Figure \ref{fig:mw}.  {\it Lower right:}
  The momentum of the swept-up shell in the M82 model, with the same
  line styles as in Figure \ref{fig:mw}. The bulk of the momentum is
  supplied by radiation; the contribution from gas pressure is
  negligible.  The early contribution of the protostellar jets is
  important in the disruption of the natal cluster gas. }
\label{fig:m82}
\end{figure*}

Tables 1 \& 2 summarize our assumed galaxy, GMC, and star cluster
properties in M82.  These are all motivated by, and reasonably
consistent with, the observations summarized above.  Our results for
the disruption of GMCs are summarized in Figure \ref{fig:m82}.  The
top two panels show the shell radius as a function of time and the
velocity as a function of radius.  The main sequence lifetime of a
$120M_\odot$ star (the dashed vertical line) is comparable to the disk
dynamical time ($R_d/v_c$, the solid vertical line). The GMC is
disrupted (reaches the Hill radius) about one disk dynamical time
after the cluster forms.  The velocity of the shell reaches higher
values than those found in our Milky Way model because of the much
larger cluster masses in M82, combined with the fact that the star
cluster radii in the two galaxies are nearly the same.  Initially, the
shell velocity is comparable to the escape velocity from the
cluster. The velocity begins to slow once the swept up mass is similar
to the mass in the cluster (at $r\sim4\pc$).

%----------------------------------------------------------------------
%----------------------------------------------------------------

What is responsible for disrupting the GMC in our M82 model? 
The lower panels of Figure \ref{fig:m82} show that the cluster gas 
on small scales is
expelled by a combination of proto-stellar jets and radiation
pressure, while the overlying GMC is disrupted primarily by radiation
pressure. The contribution from the gas pressure of the HII region is
negligible over most of the evolution: the lower right panel
shows that the H II gas pressure contribution to the momentum 
is $\sim10\%$ of that contributed by radiation when the shell radius reaches the
original size of the GMC.

\subsection{The Clump Galaxy Q2346-BX 482}

Q2346-BX 482 is a redshift $z=2.26$ disk galaxy with a disk radius of
$R_d\approx 7\kpc$ and a gas mass, as estimated from inverting the
Schmidt Law in Kennicutt (1998), of $M_g\approx3\times10^{10}M_\odot$
\citep{genzel08}. We interpret the clumps in rapidly star forming
redshift $z\sim2$ galaxies as Toomre mass GMCs, with radii
$\Rg\approx1\kpc$, and we model the giant clump in BX 482, as one of the
most extreme examples of this phenomenon.  With 
turbulent and circular velocities of $v_T\approx55\kms$ and
$v_c\approx235\kms$, respectively, we infer a disk scale height of
$H\approx1.6\kpc$, and using $\phi_G=3$ a GMC size of order $500\pc$.
For comparison, the observed clumps are modestly larger than this, 
around $1\kpc$.  

The mass of the central star cluster is set so that the luminosity
matches that observed, roughly $L\approx4\times10^{11}L_\odot$
(Genzel, private communication).  We take the radius of the star
cluster to be $\sim 7\pc$ (see \S\ref{sec:size}).  In reality, there
will likely be many star clusters, with a distribution given by
equation \ref{eq: cluster dndm}, and with a spread in age of several
to ten Myrs.  We assume solar metallicity, consistent with the
observations.  Finally, the gas mass of the clump is not known, but we
assume it is roughly the Toomre mass, $\sim10^9M_\odot$. This is
consistent with the mass of ionized gas for the observed luminosity of
the clump, at the observed size $\Rg\approx 1\kpc$, and for a stellar
population less than $4$ Myrs old (eqs.~\ref{xi} \& \ref{eq:n_L}).

\begin{figure*}
\centerline{\psfig{file=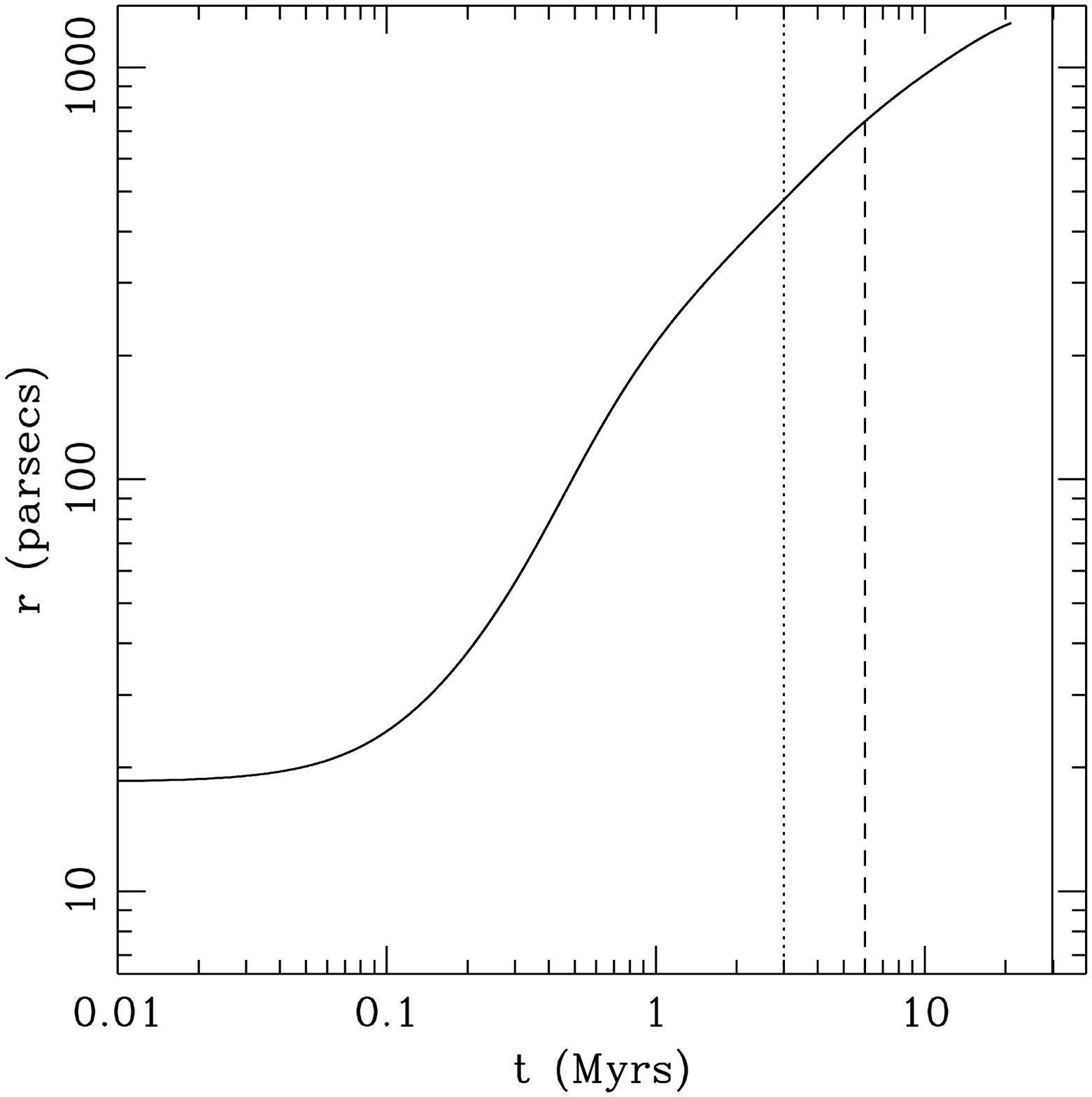,width=6cm}\psfig{file=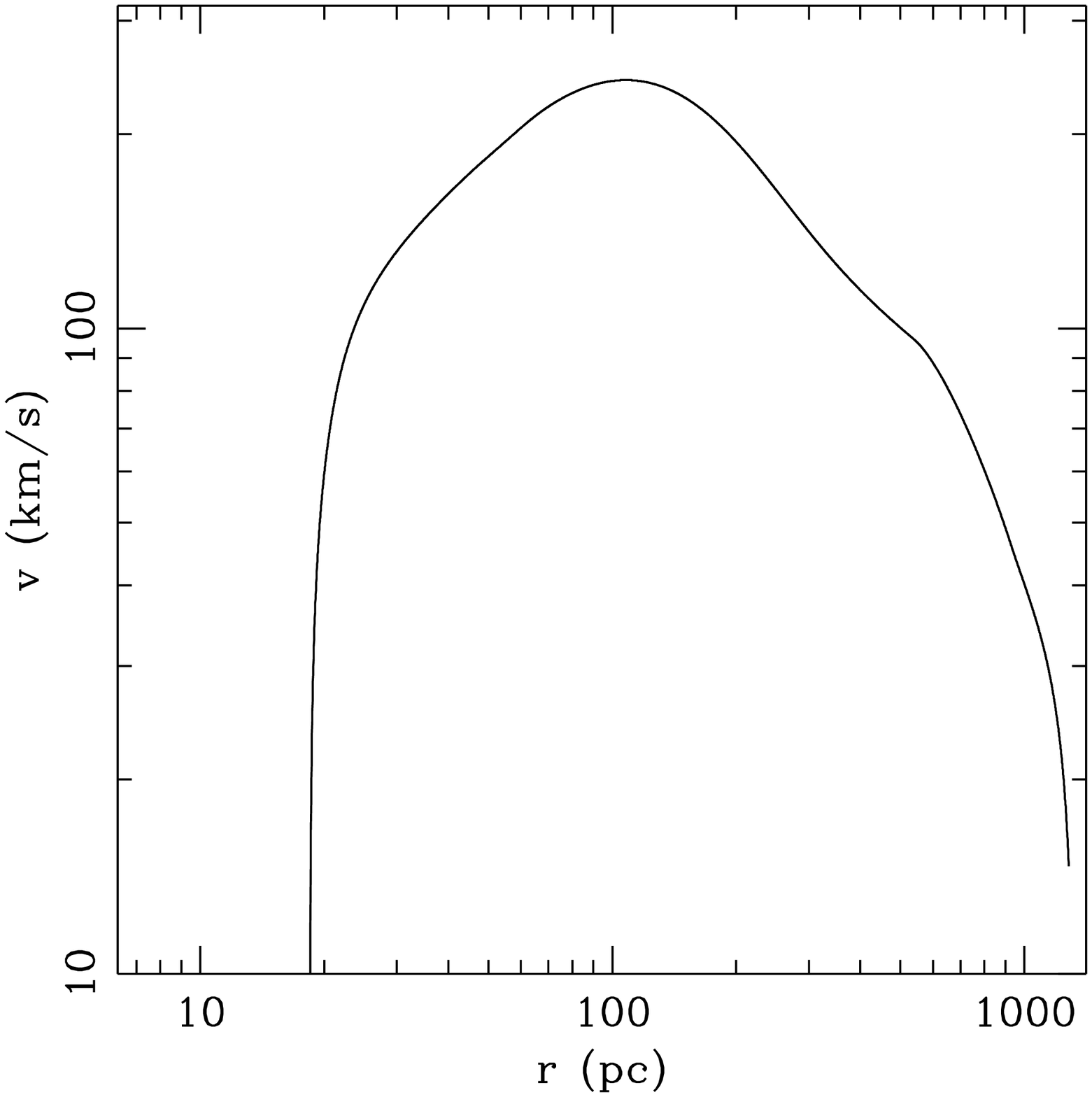,width=6cm}\psfig{file=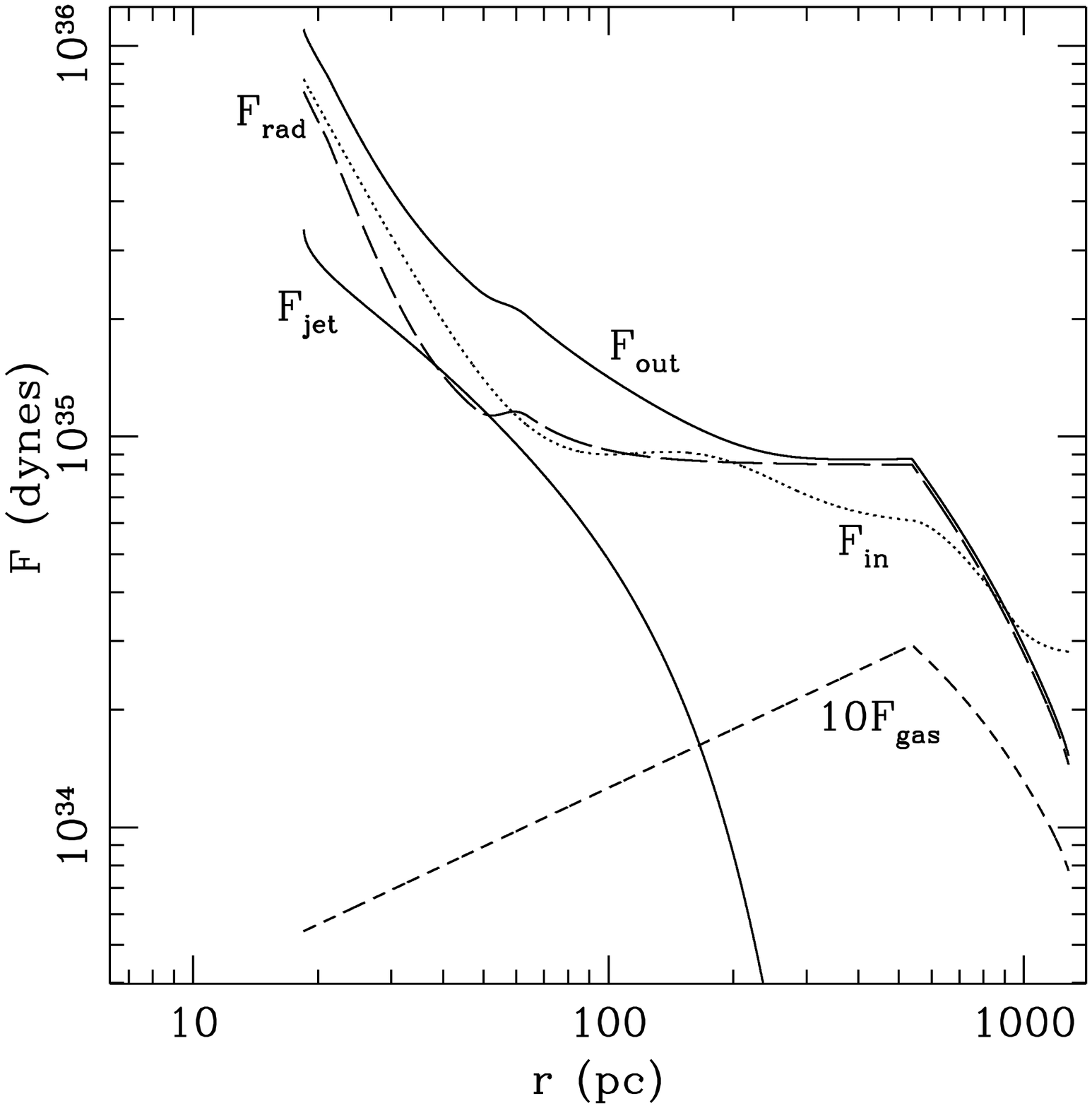,width=6cm}}
\caption{ {\it Left:} The radius of the shell as a function of time in
  the model for the giant clump in the $z = 2.26$ galaxy Q2346-BX 482.
  The dotted, dashed, and solid lines mark when the first cluster SNe
  explode, the central cluster luminosity drops to $1/3$ of its
  initial value, and $t$ reaches the disk's dynamical timescale,
  respectively.  {\it Middle:} The velocity of the swept-up shell. For
  the vast majority of the evolution, the shell velocity is comparable
  to the observed velocity dispersion of the gas, $\sim 55\kms$
  \citep{genzel08}.  {\it Right:} The forces in our model for the
  giant clump in BX482, with line styles as in Figure \ref{fig:mw}.}
\label{fig:clump}
\end{figure*}

Figure \ref{fig:clump} shows the results of our model for the giant
GMC in Q2346-BX 482.  The right panel shows that radiation pressure
dominates the evolution of the GMC at nearly all times. The GMC is
disrupted in about $15$ Myrs, half the disk dynamical time scale.  The
shell velocity, shown in the middle panel, is $\sim30-80\kms$ when the
radius is $0.5-1\kpc$, in reasonable agreement with the observed
velocity dispersion of the galaxy.  Note that the decrease in velocity
at late times (due to the decrease in radiation pressure seen in the
right panel) is probably not that physical given that the shell is in
the process of being sheared apart by the rotation of the galaxy.

Our conclusion that radiation pressure is disrupting the massive clump
in BX482 is directly supported by observations, independent of the
specific assumptions in our model: the self-gravity of the clump is
\be %$
F_{\rm grav}={GMM\over 2r^2}=3\times10^{34}
\left({M\over10^9M_\odot}\right)^2
\left({1\kpc\over\Rg}\right)^2\dynes, \ee %$
where we have scaled $M$ to the Toomre mass.  This can be compared
directly with the radiation pressure force,
\be  %$
F_{\rm rad}={L\over c}=5\times10^{34}
\left({L\over 4\times10^{11}L_\odot}\right)\dynes.
\ee  %$
The clump should thus be expanding.

%----------------------------------------------------------------------
%----------------------------------------------------------------------

\subsection{The Ultra-Luminous Infrared Galaxy Arp 220}
\label{sec:arp220}

Arp 220, at $\sim 77$\,Mpc, is the proto-typical ULIRG in the local
universe.  The gas mass of each of the $r_d\approx100\pc$ star-forming
disks in Arp 220 is $10^9M_\odot$, the circular velocity
$v_c\approx300\kms$, $v_T\approx80\kms$, and the disk scale height
$H=(v_T/v_c)r_d\approx23\pc$ \citep{downes,sakamoto}. The mean surface
density is $\Sigma_g\approx7\g\cm^{-2}$, about $100$ times larger than
that of M82 and more than a thousand times higher than in the Milky
Way. We estimate that Arp 220 has GMC masses of $\approx
5\times10^7M_\odot$, $\Rg\approx5\pc$ and a turbulent velocity in each
GMC of $\sim170\kms$, about twice that measured for the disk as a
whole.  Although the metallicity is uncertain, we take a fiducial
metallicity of $3$ times solar; this increase in metallicity is
important because it increases the dust optical depth and hence the
overall importance of radiation pressure.

\citet{wilson} found $\sim40$ young superstar clusters in and around
Arp 220; they estimate masses for about a dozen, with a number having
$\mcl\sim2-4\times 10^6M_\odot$; the largest has
$\mcl\approx10^7M_\odot$. Given the huge extinction toward the twin
disks, this is likely to be a rather conservative lower limit on the
mass of the most massive cluster in the system. The clusters are
unresolved in the HST images (which have a resolution of order $15\pc$
at the distance of Arp 220), except possibly their brightest cluster,
with a half light radius $\rcl\approx20\pc$. \citet{wilson} do not
obtain either a velocity dispersion or a half light radius for their
clusters, so they cannot calculate a dynamical mass. Rather, they use
a Salpeter IMF and \citet{bc} stellar synthesis models combined with
their photometry.

\begin{figure*}[t]
\centerline{\psfig{file=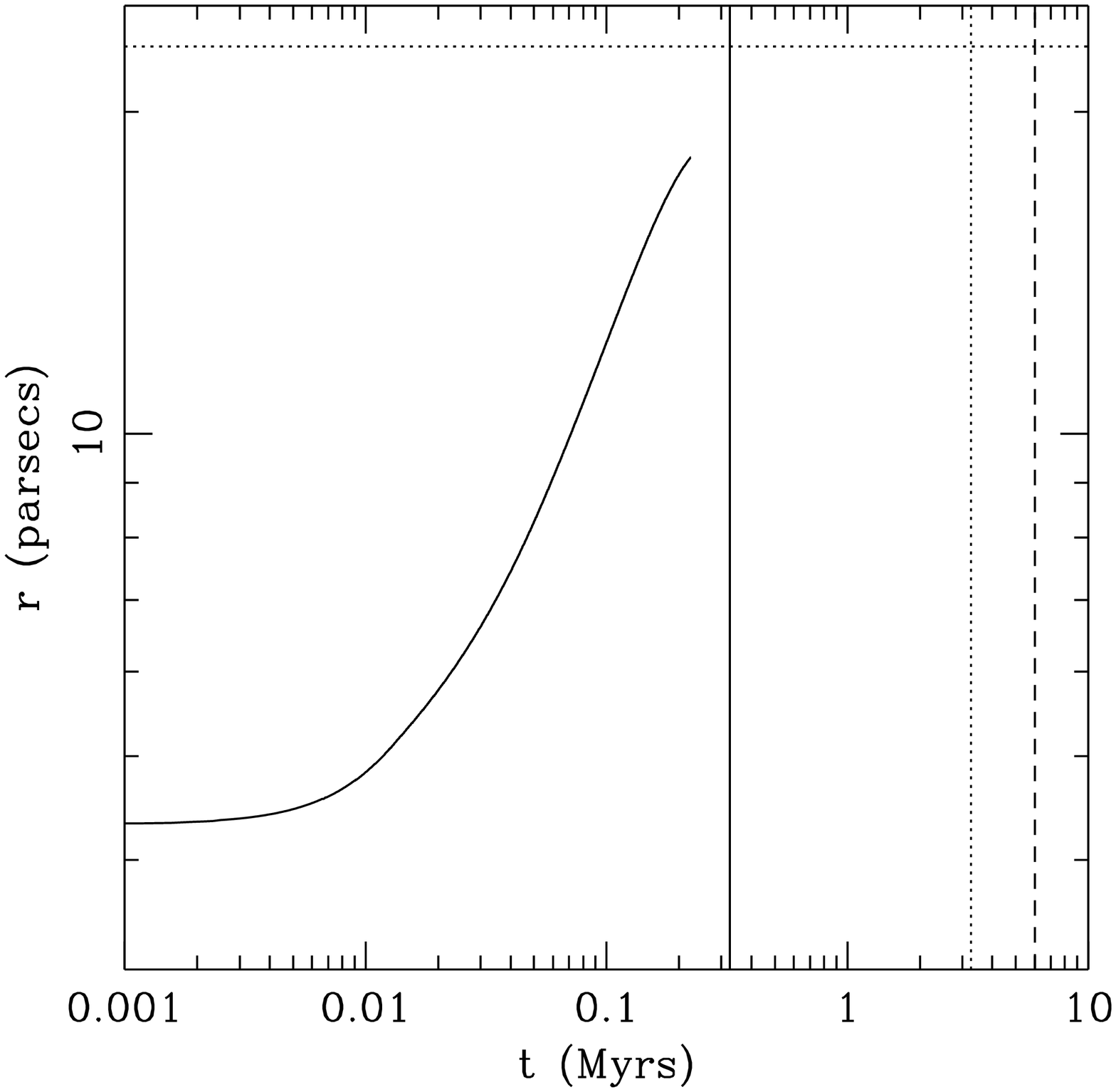,width=6cm}\psfig{file=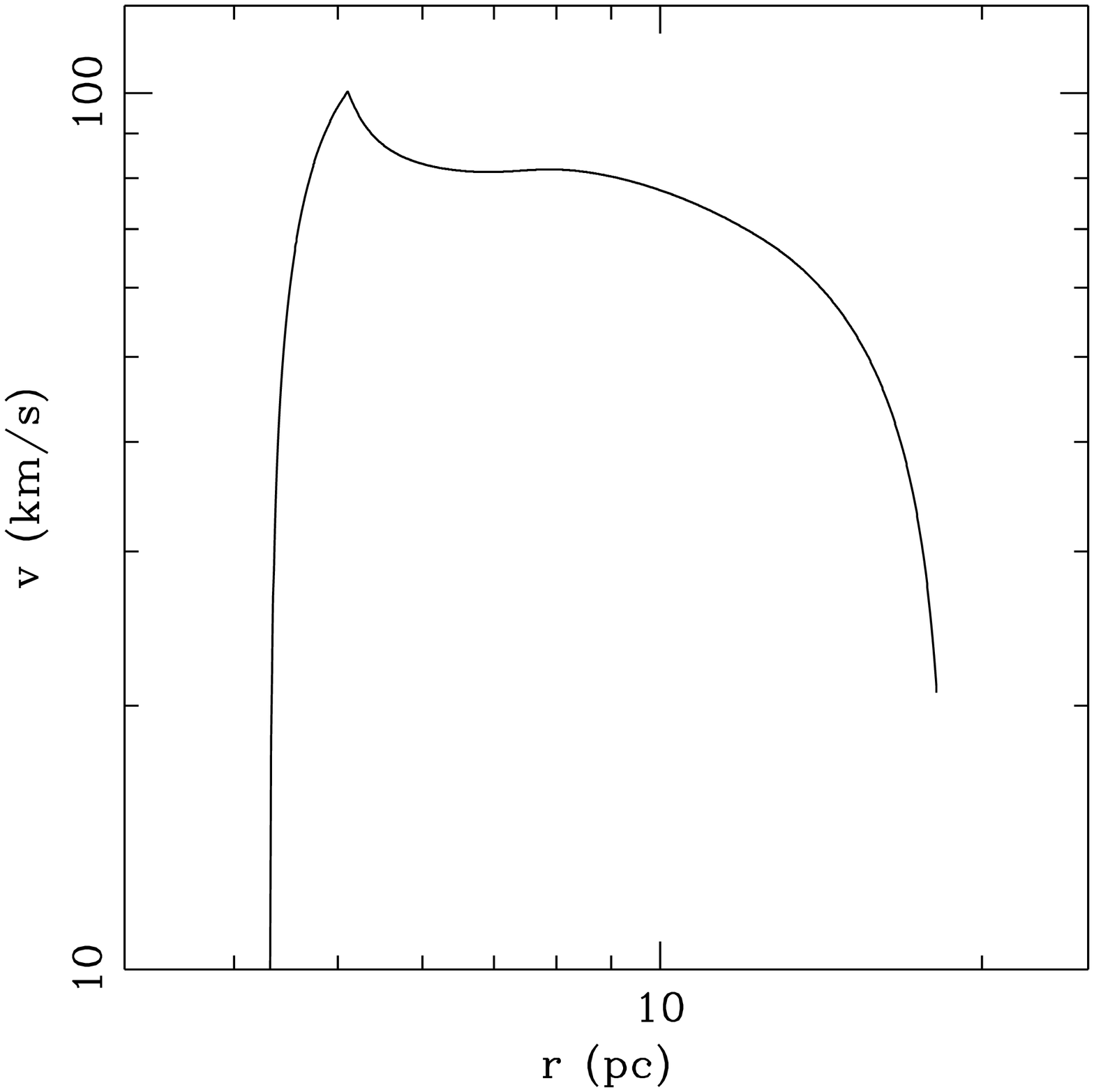,width=6cm}\psfig{file=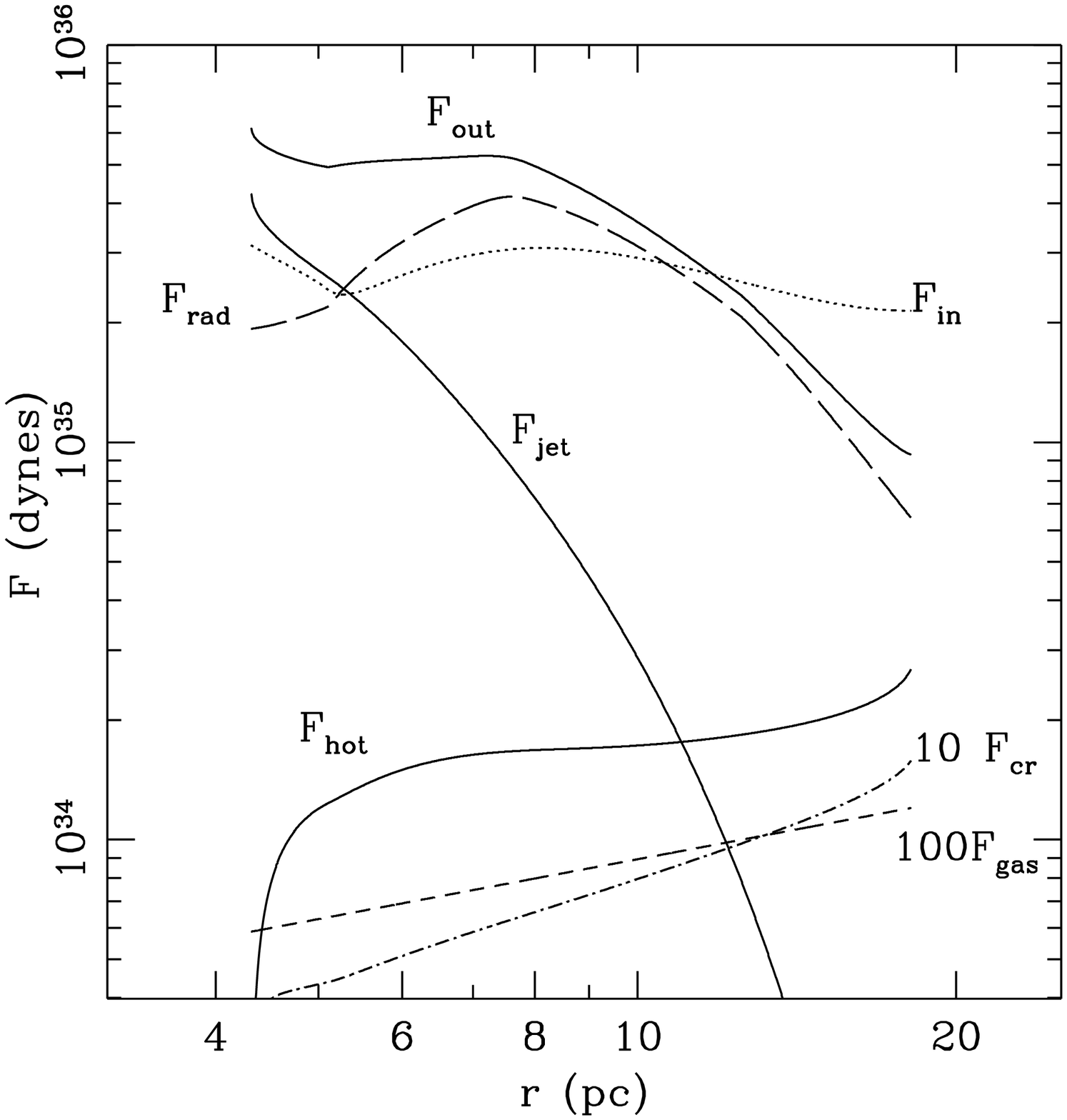,width=6cm}}
\caption{ {\it Left:} Bubble radius as a function of time in a model
  for the disruption of a GMC by a star cluster in the ULIRG Arp 220.
  The dotted, dashed, and solid lines mark when the first cluster SNe
  explode, the central cluster luminosity drops to $1/3$ of its
  initial value, and $t$ reaches the disk's dynamical timescale,
  respectively.  Note that in Arp 220, the disk and GMC dynamical
  times are short compared to the main sequence lifetime of massive
  stars, unlike in our other models (see also Table 1).  {\it Middle:}
  The velocity of the swept-up shell as a function of shell radius.
  {\it Right:} The forces as a function of the radius of the swept up
  shell. The upper most solid line is the total outward force $F_{\rm
    out}$, consisting of five components: the force exerted by
  protostellar jets (solid line; $F_{\rm jet}$), the force exerted by
  HII gas pressure {\em multiplied by 100} (short dash line; $F_{\rm
    HII}$) the force exerted by radiation pressure on dust grains
  (long dash line; $F_{\rm rad}$), the force exerted by shocked
  stellar winds ($F_{hot}$; solid), and the force exerted by
  cosmic-rays produced in stellar wind shocks ($F_{cr}$; dot-dashed).
  Radiation pressure dominates the outward force at nearly all times.
  The dotted line is the total inward force $F_{\rm in}$, dominated by
  the self-gravity of the shell.}
\label{fig:arp220}
\end{figure*}

We find that for $\mcl=1.4\times10^7M_\odot$
($L=3\times10^{10}L_\odot$), even a Toomre mass GMC
($4\times10^7M_\odot$) in Arp 220 would be disrupted (see
Fig. \ref{fig:arp220}). The disruption of the GMC occurs on the
dynamical time of the disk, well before any supernovae explode in the
GMC's central star clusters.

Our estimated GMC mass in Arp 220 ($5\times10^7M_\odot$) is a factor
ten higher than the largest GMC mass seen in M82; this is a result of
the much larger surface density in Arp 220 compared to M82.  In
contrast, the star cluster masses found so far in Arp 220 are only a
factor $2-3$ times larger than the masses of the largest clusters
observed in M82, the latter being around $2-4\times10^6M_\odot$
\citep{McCrady07}. Given that our predicted GMC star formation
efficiency is not that different in the two cases (Table 2), we expect
that more massive clusters are lurking in Arp 220. 

The right panel of Figure \ref{fig:arp220} shows the forces acting on
the shell of swept up mass in our model of Arp 220.  As in Figures
\ref{fig:mw} and \ref{fig:m82}, the force due to proto-stellar jets is
initially similar to that due to radiation pressure.  This situation
lasts only while the shell accelerates from the initial clump radius
of about 4\,pc, until the shell reaches a little less than 6\,pc. For
the rest of the evolution radiation pressure provides the only
significant outward force. The outward force supplied by ionized gas
is completely negligible; the short dash line in the Figure is the gas
pressure {\em multiplied by 100}. Both the hot gas and cosmic rays
produces by shocked stellar winds are dynamically unimportant, even
though we have assumed complete trapping of the shocked stellar wind
material (an assumption that fails in the Milky Way; Harper-Clark \&
Murray 2009).

The middle panel of 
Figure \ref{fig:arp220} shows that radiation pressure will stir the ISM
of Arp 220 to $\sim50\kms$, somewhat less than the escape velocity
from the cluster and similar to the velocity dispersion seen in CO
observations.

%--------------------------------------------------------------------------------
%----------------------------------------------------------------------

%\vspace{-0.05in}
\section{Discussion}
\label{sec:discussion}

\subsection{The Importance of Radiation Pressure}

Using four examples that cover conditions ranging from Milky Way-like
spirals to the densest starbursts (see Tables 1 \& 2), we have
explored the physical processes that can disrupt giant molecular
clouds (GMCs), one of the basic building blocks of star formation.  We
find that radiation pressure produced by the absorption and scattering
of starlight by dust grains can contribute significantly to disrupting
GMCs in nearly all types of galaxies.  By contrast, protostellar jets
are important only at early times during GMC disruption while the
thermal gas pressure in HII regions is important for GMC dispersal in
spiral galaxies like the Milky Way, but not in more luminous
starbursts.  For the Milky Way and M82, where the observations are
particularly detailed, our results demonstrate that observed massive
star clusters have precisely the luminosities and structural
properties required to disrupt Toomre-mass GMCs via radiation
pressure.

The results presented here support Thompson et al.~(2005)'s model of
radiation pressure supported star-forming galaxies.  In that paper, we
focused on the large-scale properties of star formation in galaxies
and the fueling of massive black holes in galactic nuclei.  Here we
have extended that model by taking into account the fact that star
formation is not smooth and homogeneous; rather, most stars form in
massive star clusters inside massive GMCs (\S \ref{sec:AE}).  Our
conclusion that Toomre-mass GMCs can be disrupted by radiation
pressure is qualitatively and quantitatively similar to Thompson et
al.'s conclusion that radiation pressure can regulate star formation
in galactic disks to have Toomre's $Q\approx1$.

It is useful to consider simple scaling arguments in order to
understand why, over the range of surface densities probed by the
observed Schmidt Law ($10^{-3}\,\,{\rm g \,\,cm^{-2}}\lesssim
\Sigma_g\lesssim10$\,\,g cm$^{-2}$), radiation pressure is the most
viable mechanism for GMC disruption. To rough approximation, the
self-gravity of the gas in a GMC is
\be  %$
F_{\rm sh}=\phi_G^2{GM_TM_T\over 2H^2}\propto \frac{M_T^2}{H^2}\propto M_T\Sigma_g,
\label{grav}
\ee %$
which varies by a more than a factor of $\sim10^6$ from normal
galaxies to starbursts.

We can compare this force directly to the radiation pressure force.
In the optically thin limit\footnote{By this we mean that the GMC is
  optically-thick to the UV, but optically-thin to the re-radiated FIR
  emission.}  \be F_{\rm rad}=\frac{L}{c}=\frac{\Psi M_*}{c}, \ee
where $\Psi$ is the light-to-mass ratio in cgs units.  Thus,
\begin{eqnarray}
\frac{F_{\rm sh}}{F_{\rm rad}}&\sim&\frac{\phi_G^2Gc}{2\Psi}\frac{M_T}{H^2}
\left(\frac{M_T}{M_*}\right) \nonumber\\
&\sim&1\left(\frac{3000\,\,{\rm cgs}}{\Psi}\right)
\left(\frac{0.02}{\eg}\right)\left(\frac{\Sigma_g}{2\times10^{-3}\,{\rm g\,\,cm^{-2}}}\right),
\label{thin}
\end{eqnarray}
where we have scaled to values appropriate to the Galaxy.  We see that
if $\eg$ increases with gas surface density, as our calculations
indicate (e.g., Table 2), then radiation pressure provides a plausible
mechanism for GMC disruption in both spiral galaxies like the Milky
Way and somewhat denser and more luminous galaxies.

For galaxies with sufficiently large surface densities,
$\Sigma_g\gtrsim0.5$\,g cm$^{-2}$, GMCs will be opaque to the emission
by dust in the far-infrared.  This increases the radiation pressure
force so that (in the thin shell approximation used here) \be
F_{\rm rad}=\tau_{\rm rad}\frac{L}{c}\propto
M_*\Sigma_g,\label{prad_thick} \ee where $\tau_{\rm rad}=\kappa_{\rm
 FIR}\Sigma_{\rm sh}/2$, $\kappa_{\rm FIR}$ is the Rosseland mean opacity
of the GMC in the FIR and $\Sigma_{\rm sh}$ is the surface density of
material in the shell.  Comparing the optically thick radiation
pressure force with that due to gravity, we find that
\begin{eqnarray}
\frac{F_{\rm sh}}{F_{\rm rad}}&\sim&\frac{4 \pi Gc}{\kappa_{\rm FIR}\Psi}
\left(\frac{M_T}{M_*}\right)\nonumber\\
&\sim& 1 \left(\frac{3000\,\,{\rm cgs}}{\Psi}\right)
\left(\frac{30\,\,{\rm cm^2\,\,g^{-1}}}{\kappa_{\rm FIR}}\right)
\left(\frac{0.25}{\eg}\right),
\label{thick}
\end{eqnarray}
where we have scaled to a relatively high value for the Rosseland-mean
dust opacity (see below).  Note that the ratio in equation
(\ref{thick}) does not explicitly depend on stellar/gas mass, because
both $F_{\rm sh}$ and $F_{\rm rad}$ are $\propto M^2$.  Using the
scalings in Appendix \ref{appendix:Forces}, it is easy to see that no
other stellar feedback process has this property.  Indeed, most of the
previously suggested support mechanisms scale as $F\propto M_*^\beta$
with $\beta\leq1$, viz, HII gas pressure, stellar winds, and pressures
associated with shocked stellar winds.  For this reason, although many
feedback mechanisms are competitive with gravity in normal spirals,
the self-gravity of the disk quickly overwhelms the forces due to
stellar feedback in starburst galaxies.  In contrast to these
other feedback processes, radiation
pressure in optically thick gas scales as $F_{\rm rad}\propto
M_*M_g$, so that it is at least in principle possible that radiation
pressure can disrupt GMCs even in the densest, most gas-rich
environments (e.g., ULIRGs and $z \sim 2$ galaxies).  {Radiation
pressure is, to our knowledge, the only stellar feedback process
that has this property.}

%---------------------------------------------------
\begin{figure*}[t]
\centerline{\psfig{file=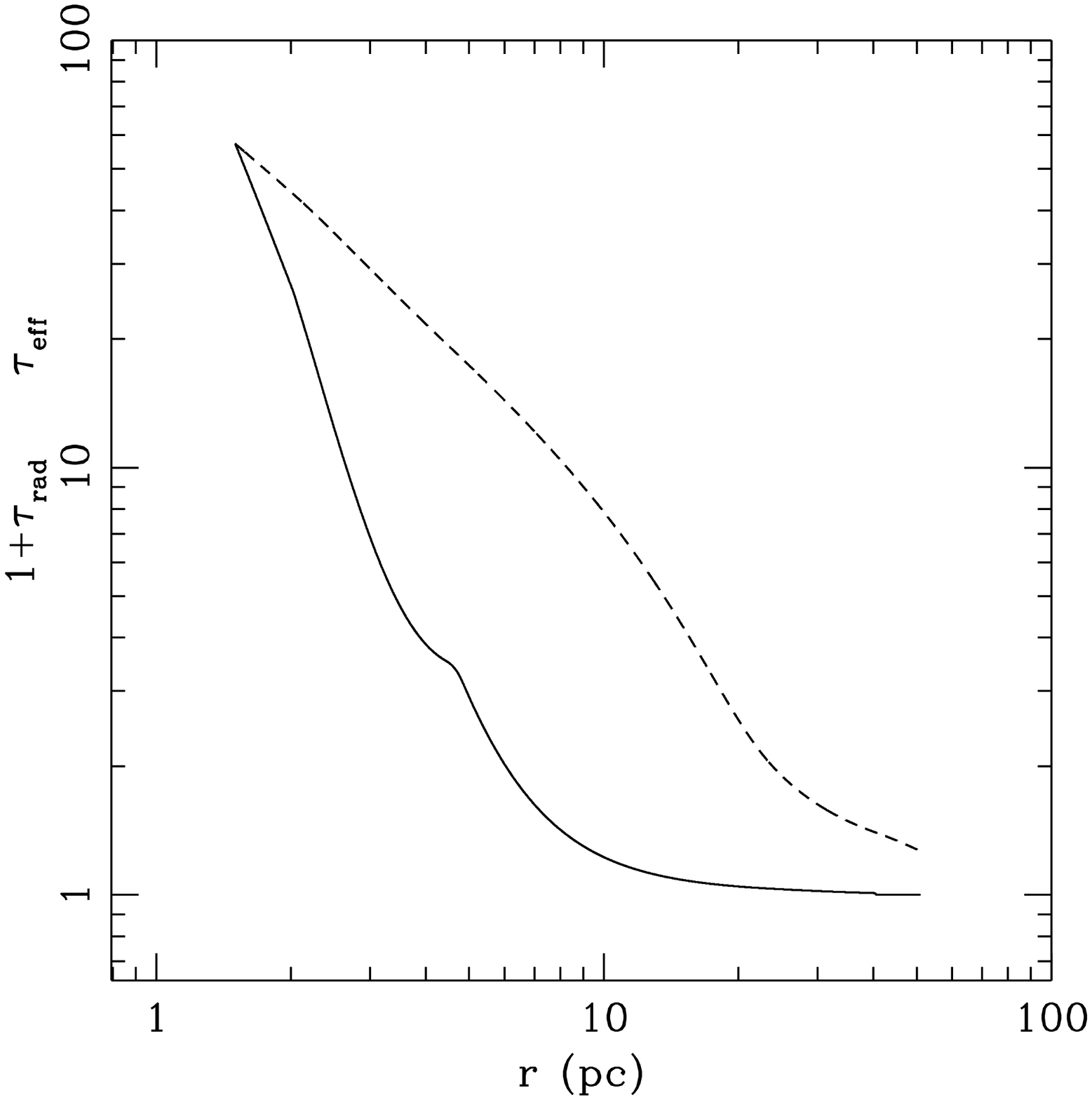,width=8.cm}\psfig{file=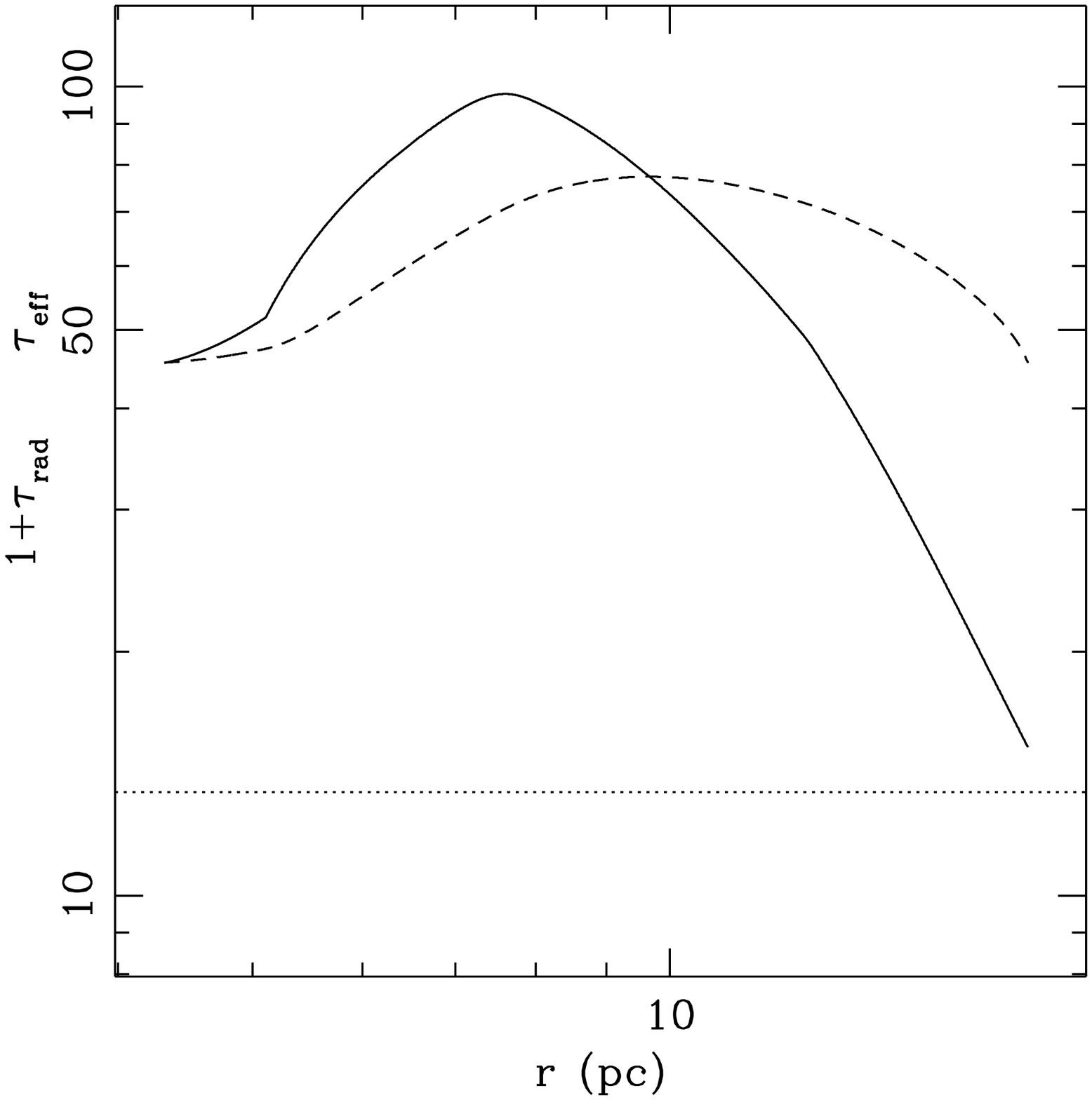,width=8.cm}}
\caption{The Rosseland mean optical depth $(1+\tau_{\rm rad})$ (solid lines) and the effective
optical depth $\tau_{\rm eff}$ (eq.~\ref{taueff}) (dashed lines) as a function of radius in the 
M82 (left) and Arp 220 (right) models, respectively.
Note that $\tau_{\rm eff}$ is a factor $\sim3$
larger than the mean optical depth of the disk in Arp 220 (the horizontal 
dotted line), enhancing the magnitude
of the radiation pressure force. }
\label{fig:tau}
\end{figure*}
%---------------------------------------------------

Equation (\ref{thick}) shows that the efficiency of star formation in
a GMC at very high densities is sensitive to the metallicity and dust
composition, which influence the FIR opacity $\kappa_{\rm FIR}$, and
to the stellar IMF, which determines the light to mass ratio of the
stellar population $\Psi$.  In very dense environments there are some
reasons for suspecting that the IMF may be top heavy (e.g.,
\citealt{murray09}), as appears to be the case in regions of massive
star formation more generally \citep{krumholz07}.  If this is indeed
the case, it would increase $\Psi$ and thus decrease the $\eg$
required for GMC disruption.\footnote{Note that as IMF becomes
  arbitrarily top-heavy $\Psi\rightarrow 4\pi Gc/\kappa_T$, where
  $\kappa_T$ is the Thomson opacity.  This sets a {\it minimum} on the
  ratio $F_{\rm sh}/F_{\rm rad}$ for any stellar population:
  $\left.F_{\rm sh}/F_{\rm rad}\right|_{\rm min}
  \rightarrow\left(\kappa_T/\kappa_{\rm FIR}\right)\eg^{-1}
  \sim10^{-2}\left(30\,{\rm cm^2\,\,g^{-1}}/\kappa_{\rm
      FIR}\right)\eg^{-1}$.}  For a relatively normal IMF, however,
the star formation efficiency in GMCs must be appreciable at high
densities, with $\eg \sim 0.25$ or perhaps even larger.

\subsection{The FIR Optical Depth}
\label{sec:tau}

A key part of our argument for the importance of radiation pressure is
the fact that star clusters have very high surface densities, which
can trap the FIR radiation produced by dust, thus enhancing the
radiation force by a factor of $\sim \tau_{\rm rad}$ in the
optically-thick limit (eq. [\ref{thick}]).  Figure \ref{fig:tau} shows
the Rosseland mean dust optical depth ($\tau_{\rm rad}$) through the
shell as a function of radius in the M82 and Arp 220 models (compare
with Figs.~\ref{fig:m82} \& \ref{fig:arp220}).  In the M82 model,
$\tau_{\rm rad}>1$ for $r\lesssim20$\,pc, while for Arp 220 $\tau_{\rm
  rad}\gg1$ at all radii.  In our Milky Way models, by contrast, we
find that the GMC is essentially always in the optically-thin limit,
i.e., opaque to the UV but not to the FIR.  Figure \ref{fig:tau} also
shows the {\it effective optical depth} in both the M82 and Arp 220
models (dashed lines), which we define as
\be  %$
\tau_{\rm eff}\equiv {P_{\rm rad}\over (Lt/c)}. \label{taueff}
\ee  %$
The effective optical depth quantifies the enhanced coupling of
photons emitted by a cluster in the center of a GMC compared to
photons originating at the mid-plane of a uniform density disk. In M82
the effective optical depth at the end of the bubble evolution is
about equal to the mean optical depth at the mid-plane.  By contrast,
in our Arp 220 models, the momentum deposited per photon in the bubble
shell is a factor of $\sim3$ larger than would deposited by a photon
traversing a uniform density disk. This shows that the effect of the
radiation pressure may be three times larger than calculated using the
mean mid-plane optical depth, as was done in \citet{thompson05}.  
This effect increases the importance of radiation pressure in the densest
galaxies, where it is needed most, effectively decreasing the
normalization in equation (\ref{thick}).  Note also that because most
of the star formation --- and thus radiation --- occurs in a few
massive star clusters in the most massive GMCs (\S\,\ref{sec:AE}),
there is very little ``cancellation'' due to different radiation
sources driving the gas in different directions (as was suggested by
\citealt{socrates2006}); the distribution of radiation sources in
galaxies is not well-approximated as infinite and homogeneous
(\S\,\ref{sec:AE}).

Given the turbulent and clumpy nature of the ISM in GMCs, one may
question whether or not the photon coupling efficiency is as large as
$\tau_{\rm rad}$ or $\tau_{\rm eff}$, since these expressions assume
uniform shells of matter.  We have, after all, used the argument that
GMCs are porous to argue that hot gas from shocked stellar winds
escapes rapidly from GMCs in the Galaxy.  The optical depth $\tau_{\rm
  rad}$ is measured from the center of the protocluster outward and is
proportional to the column density of overlying gas; the latter has
been observationally measured by a number of authors (e.g.,
\citealt{goodman}; \citealt{wong}) and is consistent with a log-normal
distribution. Numerical simulations also find log-normal surface
density distributions \citep{osg}.\footnote{The observations measure
  $\tau_{\rm rad}$ along the line of sight from the Earth through the
  cloud rather than from the center of the cloud outward, but the two
  surface density distributions should not differ dramatically.}  In
the notation of \citet{osg}, \citet{goodman} find a dispersion in the
logarithm of the column density of $0.11<\sigma<0.22$, corresponding
to a range of mean logarithmic column densities $0.01<\mu<0.05$. This
agrees well with the high Mach number turbulence simulations of
\citet{osg}.  For $\mu=0.05$, $99\%$ of sight lines have
$\tau/\bar\tau\gtrsim0.5$, where $\bar\tau$ is the (angular) mean of
the optical depth. There are thus very few optically thin sight lines
until the mean optical depth is itself of order unity.  This
conclusion is based on observations that do not probe, and simulations
that do not include, some of the physics relevant to our models (e.g.,
the Rayleigh-Taylor and photon-bubble instabilities; Blaes \& Socrates
2003; Turner et al.~2007; Thompson 2008; Krumholz et al.~2009).
Nonetheless, these results provide some support for our argument that
the dense ISM will be opaque to FIR radiation, thus significantly
enhancing the radiation pressure force in the optically thick limit.
In the absence of very large changes in the IMF ($\Psi$) or the
average dust-to-gas ratio ($\kappa_{\rm FIR}$), this enhancement is
{\it required} for radiation pressure to disrupt GMCs in dense
starbursts.   For Milky Way-like galaxies, our conclusions are not
sensitive to the radiation force in the optically thick limit, since
the GMCs/star clusters are not optically-thick except on very small
scales.

\subsection{The Origin of Large-Scale Turbulent Motions}

In all of our calculations, the shell velocity at $r\approx H_d$ is
similar to the turbulent velocity required to maintain Toomre's 
$Q \sim 1$ in the ambient disk, thus staving off self-gravity on the largest
scales. This is not a coincidence: disrupting the GMC --- which we 
have explicitly taken to be the Toomre mass --- requires a force
comparable to that needed to stir up the gas in the disk proper to $Q
\sim 1$.  We have interpreted the shell velocity at large radii as a
turbulent velocity because the disruption of the shell by shear, tidal
forces, and instabilities (e.g., Rayleigh-Taylor) will inevitably
convert this relatively ordered kinetic energy into random fluid
motions.  More sophisticated models are clearly called for, but it is
also clear that such models must include the effects of radiation
pressure, particularly in very optically thick systems like Arp 220, 
but also in models of Milky Way-like galaxies.

\subsection{The Global Star Formation Efficiency in Galaxies}

A key question we are unable to fully address is how the efficiency of
star formation in a GMC ($\eg$) relates to the global star formation
rate in the galaxy as a whole, as encapsulated by, e.g., the parameter
$\eta$ in the Kennicutt Law (eq. [\ref{eqn: Kennicutt}]) or the
observed Schmidt Law.  We briefly discuss some of the relevant issues
here, but leave a more detailed analysis of this problem to future
work.

In our models of GMCs in Milky-Way like spirals, the time for a star
cluster to disrupt the natal GMC is $1/3$ to $1/2$ the local dynamical
time. When we add to this the time for the GMC to contract to its
present size, the total time involved will be $\sim2$ longer than the
local dynamical time of the disk. If the subsequent supernovae do not
greatly prolong the process of reincorporating the bulk of the gas
back into the disk, which we suspect is correct, the time averaged 
star formation rate per unit area will be
\be  %$
\dot \Sigma_* =\eg \Sigma/(2t_{\rm dyn}) = \eta \Sigma/t_{\rm dyn},
\ee  %$
with $\eta$ given roughly by $\eta \approx \eg/2\approx 0.02$,
in good agreement with the observations.

The number of clusters capable of disrupting a Toomre mass GMC, in our
feedback model, should be proportional to the number of Toomre masses,
$\sim (R/H)^2\sim700$. The corresponding number of giant HII regions
observable at any time is $(R/H)^2 t_{\rm MS}/(2t_{\rm dyn})\approx30$,
where $t_{\rm MS}\approx3.6\times10^6$ years is the lifetime of an
early O star, which is also the time over which a star cluster will
emit a large luminosity of ionizing photons. This estimate is
consistent with the fact that half of the free-free emission in the
Galaxy is produced by about $17$ giant HII regions \citep{MR}.

The GMC star formation efficiency is $\eg=0.38$ for our fiducial Arp
220 model (Table 2); more generally, it is $\eg \sim 0.25$ in the
optically thick limit for typical IMFs and dust to gas ratios
(eq.~[\ref{thick}]). This suggests that $\eta$ is also $\approx0.25$,
higher than implied by the Kennicutt relation (eq.~[\ref{eqn:
  Kennicutt}]), although reasonably consistent with the conclusions of
\citet{bouche}.  However, this estimate assumes that after the
dispersal of the GMC, gas falls back into the disk and forms a new GMC
in a single dynamical time; as the left panel of Figure
\ref{fig:arp220} shows, the timescale for cluster disruption is in
fact $\sim1/10$ of the timescale for the cluster luminosity to drop
significantly.  Thus, the star cluster luminosity can drive motions in
the remaining gas $\sim v_T\sim(H/R)v_c$, sufficient for hydrostatic
equilibrium over many ($\sim10$) dynamical timescales.  Because the
photon diffusion timescale is rapid compared to the dynamical
timescale, the medium cannot be supported stably (Thompson et
al.~2005; Thompson 2008).  Hydrostatic balance will only be maintained
in a statistical sense within a volume $\sim 4H^3$ of the star
cluster.  After the cluster's luminosity decreases on a timescale
$\sim t_{\rm MS}$ the gas will recollapse to form a new GMC and the
process will repeat until gas exhaustion.  We suspect that this may
reduce $\eta$ by a factor of $\sim t_{\rm dyn}/t_{\rm MS}$, but
clearly more work is needed.

\acknowledgments

We thank Reinhard Genzel and Chris McKee for useful conversations.
N.M.~is supported in part by the Canada Research Chair program and by
NSERC.  E.Q.~is supported in part by NASA grant NNG06GI68G and the
David and Lucile Packard Foundation.  T.A.T.~ is supported in part by
an Alfred P.~Sloan Fellowship.

%------------------------------------------------------------------------------------------------

\begin{appendix}

\section{Forces Included in the Models} 
\label{appendix:Forces}

In this section we describe the forces acting on gas surrounding
massive star clusters embedded in giant molecular clouds.

\subsection{Inward Forces}

\subsubsection{Gravity}
We assume that the gravitational force acting on the bubble shell
consists of three components,
$F_{\rm grav}=F_{\rm stars}+F_{\rm shell}+F_{\rm disk}$. The central star cluster
exerts a force on the bubble shell given by
\be %
F_{\rm stars}=-{GM_*M_{\rm sh}\over r^2},
\ee %
the shell self-gravity is
\be %
F_{\rm shell}=-{GM_{\rm sh}^2\over 2r^2},
\ee %
while the mass in the galactic disk exerts a force
\be %
F_{\rm disk}=-{v_c^2M_{\rm sh}\over R_d}{r\over R_d}.
\ee %
The last force is that exerted on the part of the shell rising
vertically away from the disk; gas in the plane of the disk will not
feel this force, but we ignore this complication, just as we ignore
Coriolis forces.

\subsubsection{Ram Pressure}
As the shell expands into the ISM of the galaxy it will sweep up
gas. This swept up gas exerts a ram pressure on the shell giving a
force
\be %
F_{\rm ram}=-{dM_{\rm sh}\over dr}v^2,
\ee %
where $dM/dr$ is given by the appropriate derivative of
equation (\ref{eq:mass}) or (\ref{eq:mass_d}).

\subsubsection{Turbulent Pressure}
The pressure in the interstellar medium and in the GMC provides an
inward force on the shell. We refer to these pressures as turbulent
pressure, although there may be other components such as magnetic or
cosmic ray pressure. We approximate them as
\be %
F_{\rm turb}=-4\pi r^2 P_{\rm ISM} 
-{GM_{\rm GMC}^2\over \Rg^2}
\left[1-\left({r\over \Rg}\right)^2\right].
\ee %
Here we have assumed that the GMC has a Larson-like  mass
distribution. A similar expression holds for clouds with $\rho(r)\sim
1/r^2$.  The pressure of the ISM is given by
\be %
P_{\rm ISM}\approx \pi G \Sigma_d^2.
\ee %

\subsection{Outward Forces}
\subsubsection{Gas Pressure Forces}
The large luminosity $Q$ (number per second) of ionizing photons
produced by the massive stars in a young star cluster will photoionize
and heat gas in the vicinity of the cluster, raising the gas pressure
above that in the neutral gas. This hot gas will exert an outward
force on the bubble wall, with a magnitude
\be \label{eq: gas force}%
F_{\rm HII}=4\pi r^2 P_{\rm HII}
\ee %
The gas pressure is taken to be 
\be \label{eq:pressure}%
P_{\rm HII}=nkT
\ee %
where $T=8000\K$ \citep{1970A&A.....7..349G} and 
\be \label{eq:HII}%
n=\sqrt{Q\over \alpha_{\rm rec} V}.
\ee %
The volume $V=4\pi r^3/3$ and the recombination coefficient
$\alpha_{\rm rec}\approx4\times10^{-13}$. In models for clusters in the
Milky Way, the shell velocity is subsonic, so the gas pressure is
roughly constant throughout the bubble. If there is any hot gas or
cosmic rays due to, for example, shocked stellar winds, they will also
be roughly isobaric, and in pressure equilibrium with the HII gas. If
the hot gas or cosmic ray pressure is in excess of the estimate given
here, the HII gas will be confined to a fraction of the bubble
volume. However, observations of Carina and massive Milky Way and LMC
clusters suggest that the pressure is well approximated by
(\ref{eq:pressure}) and (\ref{eq:HII}) \citep{HCM}.

For large enough stellar clusters, those with at least one $35M_\odot$
star, $Q$ is proportional to $L$,
\be  %$
{L\over Q}\equiv\xi\approx 8\times10^{-11}\erg,
\label{xi}
\ee  %$
or about 3.6 Rydbergs ($1 {\rm Ryd}\approx2.2\times10^{-11}$ is the
energy required to ionize hydrogen). The HII gas pressure inside the
bubble is then
\be \label{eq: gas pressure} %$
P_{\rm HII}=5.0\times10^{-10}\nonumber
\left({L\over 10^{41}}\right)^{1/2}\left({5\pc\over r}\right)^{3/2}
\left({T\over 8000K}\right)\dynes\cm^{-2},
\ee  %$
assuming a unit filling factor for the HII gas, i.e., ignoring the hot
shocked winds. 

It is helpful to have an estimate for $n$ in terms of the
luminosity. The number density $n(L,r)$ is given by
\be  \label{eq:n_L}%$
n(L,r)=\sqrt{3L\over4\pi\alpha\xi r^3}.
\ee  %$
If the bubble is breached, so that HII gas leaks out, the pressure in
the bubble will be lower than the estimate given here, but the force
experienced by the bubble wall will actually be larger. The reason is
that fewer ionizing photons are absorbed in the bubble cavity, leaving
more to heat the gas on the interior of the bubble wall. The heated
gas escapes away from the wall into a partial vacuum, exerting a force
\be %$
F=\dot M c_{\rm HII}
\ee %$
on the wall. The mass loss rate is given by $4\pi r^2 m_p n c_{\rm
  II}$.  In our numerical work we have assumed that the
HII gas leaks out of the bubble, since the rapidly expanding
ionized gas at the ionization front will exert a larger force, and we
want an upper limit on the efficacy of HII gas pressure.

In models for ULIRGs like Arp 220, the bubble expansion velocity is
larger than the sound speed of $8000\K$ gas, so the HII pressure may
be twice that in the subsonic case even if no gas escapes from the
bubble. However, the pressure associated with HII gas is negligible in
ULIRGs, as we show in the main text.

\subsubsection{Forces Associated With Shocked Stellar Winds}

Hot gas from shocked stellar winds is often thought to be important in
the formation of bubbles around massive stars or star clusters. We
argue here that it is not; see also \citet{HCM}. We further argue that
cosmic ray pressure cannot be important in Milky Way bubbles; if they
were, the bubbles would expand more rapidly than is observed. We show
in this paper that neither hot gas nor cosmic ray pressure is not
relevant in ULIRGs.

\subsubsection{The force due to hot, shocked stellar wind gas: X-Ray Constraints}
O stars emit high velocity massive winds; in clusters these winds are
seen to shock and emit diffuse x-rays at $\sim \KeV$ energies
\citep{Seward,Oey96,SSN05}.  If the stellar winds are confined to the
bubble interior, the associated pressure can be far larger than the
ram pressure of the wind \citep{1975ApJ...200L.107C}. The force is
given by
\be  %$
F_h = 4\pi r^2P_h,
\ee  %$
where  $P_h = 2E_h/(3V)$. The energy equation for the hot gas is
\be \label{eq: energy} %$
{dE_h\over dt} = L_w - 4\pi r^2 P_h v_{\rm sh} - \Lambda n_h^2 V;
\ee  %$
recall that $V$ is the volume of a bubble of radius $r$ and $\Lambda$
is the cooling function. The wind luminosity
$L_w=(1/2)\eta(v_\infty/c)L_{\rm bol}$,
($\eta\approx0.5$ is the fraction of stellar luminosity scattered in
the wind) and $L_{\rm bol}$ is the bolometric luminosity of the cluster.

We follow \citet{1975ApJ...200L.107C} to find the number density $n_h$
of hot gas, i.e., we assume that heat conduction (at the Spitzer rate
$C_{\rm Sp} T_h^{5/2}$) drives a mass flow into the bubble interior
(possibly supplemented from cold gas clouds in the interior of the
bubble) at a rate
\be  \label{eq: mass} %$
{dM_h\over dt}= {16\pi m_P C_{\rm Sp} T_h^{5/2}r\over 25 k}.
\ee  %$
Given $E_h$ and the radius of the bubble we find the hot gas pressure;
from equation (\ref{eq: mass}) we find $n_h$, and hence the
temperature. With $n_h$ and $T_h$, we can find the cooling rate (the
third term on the right hand side of the energy equation (\ref{eq:
  energy}).

The solar-mass stars in clusters also emit x-rays; in low resolution
(non-Chandra) observations, these stars appear as diffuse
emission, so care must be taken to distinguish the two.
Assuming that one can extract the stellar emission, the
diffuse x-ray flux can be used to constrain the pressure of any hot
gas component. \citet{HCM} use this to show that for clusters in the
Milky Way and the LMC, the x-ray gas is in approximate pressure
equilibrium with the HII gas, and that the HII gas has a filling
factor near unity (greater than $\sim0.1$). The implication is that
the shocked winds either radiate away the bulk of their energy, or
else leak out of the bubble wall. In either case, they do not play any
role in the dynamics of the bubble, a conclusion also reached by
\citet{McKee84}.

\citet{HCM} show that observations of ultracompact HII regions
\citep{Rauw02,tsujimoto} reveal a similar story.

If the bubbles that form around star clusters in ULIRGs are also
leaky, the shocked wind pressure would be negligible, as in the Milky
Way. However, even if we assume that the winds are perfectly trapped,
the high ambient pressures in ULIRGs ensure that the shocked gas is
not dynamically important.

To see this, note that we expect a total number of young (ionizing)
clusters in a ULIRG to be
\be  %$
N_{\rm cl}=\left({R\over H}\right)^2 \approx 20.
\ee  %$
If there are $\sim20$ clusters powering Arp 220, each must have
$L_{\rm cl}\approx5\times10^{10}L_\odot$, and a mass
$\mcl\approx2\times10^7M_\odot$ assuming a normal
\citep{muench02} IMF; this cluster luminosity is also that
needed to disrupt a GMC in Arp 220, as we showed in \S
\ref{sec:arp220}. The wind luminosity of a single such cluster is
$L_w\approx 7\times10^{41}\ergs$. We note that
$L_{\rm cl}\approx5\times10^{10}$ is about $3-4$ times the luminosities of
known young star clusters in Arp 220 \citep{wilson}.

From Figure \ref{fig:arp220} the bubble disrupts the GMC
($\Rg\approx10\pc$) after a time $t\sim 5\times10^{12}\s$. Assuming no
losses, the wind energy accumulated in the bubble is
$E_h=2.5\times10^{54}\erg$. The pressure is
$2E_h/3V=10^{-5}\dynes\cm^{-2}$, about equal to the ambient pressure
in the disk but well below the radiation pressure, $\tau
F/c\approx10^{-4}\dynes\cm^{-2}$.

The force we have estimated from the hot gas is
$1.6\times10^{35}\dynes$; in the right panel of Figure \ref{fig:arp220} the force from
the hot gas is a factor of $\sim5$ below this estimate. The reason is
that the hot gas suffers radiative losses, so that we have
overestimated $E_h$. We note that in our numerical work we have
assumed that the hot gas fills the volume of the bubble; if it does
not, the cooling rate will be higher than we have calculated.

\subsection{The force due to wind generated cosmic rays}
In addition to producing hot gas, wind shocks will generate cosmic
rays. Perhaps a third of the wind energy may be expected to be
deposited into cosmic rays. If, as observed in the Milky Way, the hot
gas is advected out of the shell, it will probably take the cosmic
rays with it, so that both the hot gas and the cosmic rays are not
dynamically relevant. 

We note that the pressure due to cosmic rays, if perfectly confined,
would be $1/6$ that of similarly confined shocked winds. As noted in
the introduction, the dynamics of bubbles in the Milky Way and the LMC
are not consistent with the high pressures associated with trapped hot
gas; it follows that cosmic rays are probably also not confined to the
interiors of such bubbles.

If the shocked winds in ULIRGs also flow through the swept up shell,
cosmic rays are unlikely to be dynamically important. Even if the
cosmic rays are confined, the cosmic ray pressure is still
negligible. The cluster bolometric luminosity is
$2\times10^{44}\ergs$, so the wind luminosity is
$5\times10^{41}\ergs$. Assuming $1/3$ of this is deposited in cosmic
rays, the cosmic ray luminosity is
$L_{\rm cr}\approx1.6\times10^{41}\ergs$. The cosmic ray pressure when the
bubble radius is $10\pc$ is
\be  %$
P_{\rm cr}={L_{\rm cr}t_{\rm dyn}\over V}\approx7\times10^{-6}
\left({\pc\over \Rg}\right)^3\dynes\cm^{-2}.
\label{pcr}
\ee  %$
This is about a half of the mean dynamical pressure in the galaxy, but
only several percent of the dynamical pressure in the bubble.
This is still likely to be an overestimate of the cosmic ray pressure, as
cosmic rays will quickly diffuse out of the bubble.  The cosmic ray
mean free path in the Milky Way is $0.1\pc < \lambda_{\rm cr} < 1\pc$
\citep{2002cra..book.....S}; scaling to the lower value, the time for
a cosmic ray to diffuse out of a GMC in Arp 220 is
\be  %$
t_{\rm cr}={\Rg^2\over D_{\rm cr}}\approx2\times10^{11}
\left({0.1\pc\over\lambda_{\rm cr}}\right)\s,
\label{tdiff}
\ee  %$
where $D_{\rm cr}=\lambda_{\rm cr}c/3$ is the cosmic ray diffusion
coefficient. This diffusion time is $\sim10-20$ times shorter than the bubble
dynamical time, so even if the mean free path for cosmic rays in Arp
220 is substantially smaller than in the Milky Way, cosmic rays will
diffuse out of the bubbles in the ULIRG.
Figure \ref{fig:arp220} shows that the cosmic ray pressure is a
little less than one percent of the radiation pressure; the lower
cosmic ray pressure is the result of taking $P_{\rm cr}=L_{\rm cr}t_{\rm cr}/V$,
which accounts for diffusive losses, instead of equation (\ref{pcr}).
Note that we have optimistically ignored cosmic ray losses due to inelastic
proton-proton collisions, which cool the cosmic ray population
with energies $\gtrsim$\,GeV on a timescale $t_{\rm pp}
\approx 5\times10^3(10^4{\rm cm^{-3}}/n)$\,yr, where we have
scaled to the average volumetric gas density of Arp 220.

\subsubsection{Protostellar Jets \label{appendix: jets}}

While stars are actively accreting in the protocluster, we assume that
they will emit high velocity outflows or jets. Once the cluster
disrupts, the accretion halts, and the jets turn off. We assume that
this happens over a time given by $\phi_{\rm ff}$ times 
$t_{\rm ff}$ of the cluster; in addition, we allow for an extra factor
of two to account for the fact that the accretion disks will not
vanish once the cluster gas is dissipated, but instead will accrete
onto the star in a disk viscous time:
\be %
t_{\rm jet}=2\phi_{\rm ff}t_{\rm ff}.
\ee %
We take $\phi_{\rm ff}=3$.  The jets deposit momentum into the
surrounding gas at a rate
\be %
F_{\rm jet}=\epsilon_{\rm jet}\dot M_{\rm acc} v_{\rm jet}\exp^{-t/t_{\rm jet}}
\ee %
The jets expel mass at a fraction $\epsilon_{\rm jet}\approx0.1-0.3$ of
the mass accretion rate \citep{matzner_mckee}, at a velocity $v_{\rm jet}$
that depends on the mass of the star. We estimate $v_{\rm jet}$ by
calculating the escape velocity from the surface of a star of mass
$m$, using the radius of a star of that mass found from the main
sequence radius as given by the Padova models; this is a slight over
estimate of the escape velocity, since accreting stars are larger than
main sequence stars, but we compensate for this by using the escape
velocity rather than the (larger) terminal velocity of a wind or
jet. We then average over the initial mass function to yield
$v_{\rm jet}$. For a Muench et al. IMF we find $v_{\rm jet}\approx 3\times
10^7\cm\s^{-1}$, but for top heavy IMFs this can increase to
$10^8\cm\s^{-1}$. We varied $\epsilon_{\rm jet}$ over the range $0.1-0.3$
giving an effective velocity in the range $30-90\kms$, bracketing that
found by, e.g., \citet{matzner_mckee} and \citet{nakamura}. The
results did not depend strongly on this parameter, but because the
clusters we consider are so dense, we expect a substantial amount of
momentum cancellation (from jets colliding nearly head on) we used low
values for $\epsilon_{\rm jet}$ in the runs presented here.

\subsubsection{Radiation Force}

The radiation force is given by
\be \label{eqn: radiation force}%
F_{\rm rad}=(1+\tau_{\rm rad}){L\over c},
\ee %
where $\tau_{\rm rad}$ is the Rosseland mean opacity through the
shell. The gas in the shell has a temperature that is of order
$100\K$, so for most clusters in the Milky Way $\tau_{\rm rad}<1$. We
calculate the optical depth as follows: we assume the shell has a
thickness $1/10$ of its radius, and divide it into 100 pieces. We
assume that the density is constant through the shell. We calculate
$T_{\rm eff}$ from $L$ and the radius $r$ of the shell. Given $T$ and
$\rho$, we use the opacity tables of \cite{2003A&A...410..611S} to
find the Rosseland mean opacity in the outermost slice of the
shell. Subsequently we find the optical depth, then integrate the
radiative transfer equations inward through the shell.  The factor of
unity in equation (\ref{eqn: radiation force}) accounts for the fact
that the temperature of the radiation field is of order 30,000K at the
inside edge of the shell, before the photons have encountered any dust
grains; upon striking a dust grain, the photons provide an impulse
per unit time given by $L/c$.

\end{appendix}

\end{document}